%% file: sample-acmsmall.tex
\documentclass[acmsmall]{acmart}

\usepackage[utf8]{inputenc}
\usepackage[english]{babel}
\usepackage{algorithm}
\usepackage{algpseudocode}
\usepackage{textcomp}
\usepackage{xcolor}
\usepackage{xspace}
\usepackage{booktabs}
\usepackage{tabularx}
\usepackage{colortbl}
\usepackage{pgfplots}
\usepackage{adjustbox}
\usepackage{xspace}
\usepackage{listings}
\usepackage{adjustbox}
\usepackage{enumerate}
\usepackage{makecell}
\usepackage{threeparttable}
\usepackage{pifont}
\pgfplotsset{compat=1.16}
\definecolor{headerColor}{rgb}{0.8, 0.8, 0.8}

\usepackage{array} 

\usepackage{multirow}
\usepackage{enumitem}
\usepackage[most]{tcolorbox} 
\usepackage{tikz}
\usetikzlibrary{calc,fadings,decorations.pathreplacing}
\usetikzlibrary{shapes,arrows}
\usepackage{float}
\usepackage{graphicx}
\usepackage{float}
\usepackage{subcaption}
\usepackage{overpic}
\usepackage{listings}
\usepackage{hyperref}
\usepackage{url}
\hypersetup{hidelinks=true} 

\newcommand{\rA}[1]{#1}
\newcommand{\rB}[1]{#1}
\newcommand{\rC}[1]{#1}
\newcommand{\rM}[1]{#1}

\newcommand{\etal}{\hbox{\emph{et al.}}\xspace}
\newcommand{\eg}{\hbox{\emph{e.g.}}\xspace}
\newcommand{\ie}{\hbox{\emph{i.e.}}\xspace}

\newcommand{\wrt}{\hbox{\emph{w.r.t.}}\xspace}

\definecolor{darkgreen}{RGB}{0,180,0}

\newcommand*\emptycirc[1][1ex]{\tikz\draw (0,0) circle (#1);} 
\newcommand*\halfcirc[1][1ex]{%
	\begin{tikzpicture}
	\draw[fill] (0,0)-- (90:#1) arc (90:270:#1) -- cycle ;
	\draw (0,0) circle (#1);
	\end{tikzpicture}}

\newcolumntype{Y}[1]{>{\raggedright\arraybackslash}m{#1}}

\AtBeginDocument{%
  }

 \setcopyright{none}
 \renewcommand\footnotetextcopyrightpermission[1]{} 

\begin{document}

\title{An Empirical Study of False Negatives and Positives of Static Code Analyzers From the Perspective of Historical Issues}

\author{Han Cui}
\author{Menglei Xie}
\author{Ting Su}
\affiliation{%
  \institution{East China Normal University}
  \country{China}
}

\author{Chengyu Zhang}
\affiliation{%
  \institution{ETH Zurich}
  \country{Switzerland}
}

\author{Shin Hwei Tan}
\affiliation{%
  \institution{Concordia University}
  \country{Canada}
}

\renewcommand{\shortauthors}{Cui et al.}

\input{Sections/Abstract.tex}


\maketitle

\input{Sections/Introduction.tex}

\input{Sections/Methodology.tex}

\input{Sections/RQ1.tex}

\input{Sections/RQ2.tex}

\input{Sections/implication_and_discussion.tex}

\input{Sections/Experiment.tex}

\input{Sections/threats_to_validity.tex}

\input{Sections/Relatedwork.tex}

\input{Sections/Conclusion.tex}

\bibliographystyle{ACM-Reference-Format}
\bibliography{references}

\end{document}

%% file: Sections/Abstract.tex
\begin{abstract}

    Static code analyzers are widely used to help find program flaws. However, in practice the effectiveness and usability of such analyzers is affected by the problems of false negatives (FNs) and false positives (FPs).
    \rM{This paper aims to investigate the FNs and FPs of such analyzers from a \emph{new} perspective, \ie, examining the historical issues of FNs and FPs of these analyzers reported by the maintainers, users and researchers in their issue repositories --- each of these issues manifested as a FN or FP of these analyzers in the history and has already been confirmed and fixed by the analyzers' developers.}
    To this end, we conduct the \textit{first} systematic study on a broad range of 350 historical issues of FNs/FPs from three popular static code analyzers  (\ie, \textsc{PMD},  \textsc{SpotBugs}, and \textsc{SonarQube}). All these issues have been confirmed and fixed by the developers.
    We investigated these issues' root causes and the characteristics of the corresponding issue-triggering programs. It reveals several new interesting findings and implications on mitigating FNs and FPs. Furthermore,
    guided by some findings of our study, we designed a metamorphic testing strategy to find FNs and FPs. This strategy successfully found 14 new issues of FNs/FPs, 11 of which have been confirmed and 9 have already been fixed by the developers.
    Our further manual investigation of the studied analyzers revealed one rule specification issue and additional four FNs/FPs due to the weaknesses of the implemented static analysis.
    We have made all the artifacts (datasets and tools) publicly available at \rC{\textit{\url{https://zenodo.org/doi/10.5281/zenodo.11525129
            }}}.

\end{abstract}

%% file: Sections/Introduction.tex
\section{Introduction}

Static code analyzers (\emph{or} static checkers)~\cite{static_code_analysis} are commonly used to help find program flaws at the early stages of software development (after code compilation and before testing)~\cite{Afewbillion,SpongeBugs:Automatically,Warning-IntroducingCommits,howdevlopers}.
These analyzers usually target various types of program flaws including
best practice violations, code design issues, common programming mistakes and security vulnerabilities~\cite{extendedstaticcheckforjava,isthereagoldenfeature,howdevlopers}.
Otherwise, these flaws might be costly and difficult to find by manual code reviews or testing~\cite{developertestinginthe,whydontsoftwared,Exploitingcodeknowledgegraph}.
Typically, to find these flaws, these analyzers implement a number of \emph{checking rules} (or \emph{rules} for short), supported by some forms of static analysis with different complexities (\eg, syntactic pattern matching, control-/data-flow analysis, symbolic execution).

In practice, it is known that the effectiveness and usability of these analyzers could be affected by the problems of \emph{false negatives} (\emph{FN}s for short), \ie, missing \emph{true} program flaws~\cite{howmanyof,towhatextent,towhatextent_journal}, and \emph{false positives} (\emph{FP}s for short), \ie, the reported warnings are \emph{spurious}~\cite{effectiveness_2,whydontsoftwared,AreSonarQubeRules}.
\rB{To investigate FNs and FPs, most of existing studies~\cite{effectiveness_2,towhatextent,towhatextent_journal,howmanyof,AreSonarQubeRules,TomassiR21,LippBP22,MehrpourL23,LiuCFLLXNLC23,LiCFFLLLC23} evaluate the fault detection abilities or the usability of these analyzers against known faults.}
For example, Habib \etal~\cite{howmanyof} use the faults of \textsc{Defects4J} to
investigate FNs and find that these analyzers could \emph{only} find 4.5\% of the field defects; Wedyan \etal~\cite{effectiveness_2} use the known coding faults from some open-source projects to investigate FPs and find that 96\% of the warnings reported by these analyzers are spurious.
The main goal of such studies is to assess the effectiveness and usability of the evaluated analyzers in terms of \emph{general} FNs and FPs. For example, these studies find that most field defects are missed because these defects are not targeted by existing rules in the analyzers or these defects are domain-specific errors~\cite{towhatextent,towhatextent_journal,howmanyof,LiuCFLLXNLC23,LiCFFLLLC23}.

Different from these prior studies, this paper aims to investigate the FNs and FPs from a new perspective, \ie, examining the \emph{historical}, \emph{fixed} issues of FNs and FPs of these analyzers in their own issue repositories --- each of these issues manifested as a FN or FP of these analyzers and has been confirmed and fixed by developers. 
Exploring this perspective has one key benefit --- we can inspect the fixing patches and the implementations (which however the prior studies usually do not care about) of the analyzers to obtain the fine-grained insights on why FNs and FPs are induced.
These insights could be useful for the analyzers' developers to mitigate FNs/FPs at the root.
To our knowledge, \emph{no} prior work \emph{systematically} studies such issues.

To fill the gap, we studied a broad range of 350 historical issues
(corresponding to 80 FNs and 270 FPs), which were collected from the issue repositories of three popular, open-source static code analyzers (\ie, \textsc{PMD}~\cite{pmd}, \textsc{SpotBugs}~\cite{spotbugs}, and \textsc{SonarQube}~\cite{SonarQube}).
All these issues are valid and representative because they have been confirmed \emph{and} already fixed by the developers.
Specifically, to obtain a general and in-depth understanding of FNs and FPs, we study how these issues are induced (\ie, \emph{root causes}), and which characteristics of input programs could lead to these issues (\ie, \emph{input characteristics}).
To our knowledge, this is the \emph{first} systematic study to investigate the FNs and FPs of static code analyzers from the perspective of historical issues.
We will discuss and compare with the relevant work~\cite{wang2022find,Statfier} in detail in Section~\ref{sec:related_work}.

Our study aims to answer the the following three research questions:
\begin{itemize}[leftmargin=*]
      \item \textbf{RQ1 (Root Causes)}: What are the common root causes of these historical issues of FNs and FPs affecting the static code analyzers? \rA{We aim to investigate the root cause by examining the analyzers' documentations (\eg, rule specifications), the issue reports, the issue-triggering programs and the fixing patches (detailed in Section~\ref{sec:rootcause}). This RQ identifies the reasons why the FNs/FPs are induced and helps developers avoid or mitigate such issues.} 
      \item \textbf{RQ2 (Input Characteristics)}: What are the characteristics of input programs leading to these historical issues of FNs and FPs of the static code analyzers? \rA{We aim to investigate the input characteristics by examining the analyzers' documentations (\eg, rule specifications), the issue reports and the issue-triggering programs and their representations (detailed in Section~\ref{sec:characteristic}). This RQ identifies which characteristics of the input programs are pivot for inducing FNs or FPs and helps developers design better test programs or testing strategies for static code analyzers.}
\end{itemize}

Answering these two questions is beneficial to both the analyzers' developers and the researchers in this field.
The answers can provide new insights on how to improve these analyzers (\eg, mitigating or unveiling FNs and FPs), thus complementing those general studies on only evaluating the effectiveness or the usability of these analyzers~\cite{effectiveness_2,towhatextent,towhatextent_journal,howmanyof,AreSonarQubeRules,TomassiR21,LippBP22,MehrpourL23,LiuCFLLXNLC23,LiCFFLLLC23}.
Through answering \textbf{RQ1} and \textbf{RQ2}, we obtained several new interesting findings and implications that shed light on mitigating FNs/FPs (Section~\ref{sec:implication}). Therefore, we aim to investigate the usefulness of these findings:
\begin{itemize}[leftmargin=*]
	\rM{
		\item \textbf{RQ3 (Usefulness of Our Findings)}:
		How can some of our study's findings from \textbf{RQ1} and \textbf{RQ2} help identify issues of FNs and FPs in these static code analyzers? We aim to show some proof-of-concept demonstrations (in Section~\ref{sec:proof_of_concept}) to validate the usefulness of our study's findings.
	}
\end{itemize}
Specifically, we designed a metamorphic testing strategy to automatically find FNs/FPs (Section~\ref{sec:transformation}). This strategy helped us find 14 new issues (12 FNs and 2 FPs) of the studied analyzers, 11 of which have been confirmed and 9 have been fixed. Additionally, we further manually examined the implementations of some rules of the studied analyzers (Section~\ref{sec:analyzer_weaknesses}). We found one rule specification issue and additional four FNs/FPs due to the weaknesses of the implemented static analysis. We have reported these issues to the developers and all have been confirmed/fixed.

In summary, our work has made the following contributions:

\begin{enumerate}[label=\textbullet,leftmargin=*]
      \item We conduct the \emph{first} systematic study to investigate FNs and FPs of static code analyzers from the new perspective of the historical issues. We construct a dataset of 350 historical issues of FNs and FPs from the studied analyzers to serve as the basis of our study and future research in this field.

      \item We study the 350 historical issues of FNs and FPs to investigate their root causes and input characteristics and identify several new interesting findings.

      \item We discuss the implications of our study to shed light on mitigating FNs and FPs. We also demonstrate the usefulness of our study's findings by two proof-of-concept demonstrations.

      \item We have made all the artifacts (datasets and tools) publicly available at \rC{\textit{\url{https://zenodo.org/doi/10.5281/zenodo.11525129}}}.
\end{enumerate}

%% file: Sections/Methodology.tex
\section{Study Methodology}
\label{sec:metho}
This section details the methodology of our study. Specifically,
\rC{Section~\ref{sec:background} gives some background knowledge on the static code analyzers},
Section~\ref{sec:studied_static_analyzers} introduces the selected analyzers for our study,
Section~\ref{sec:issue_collection} presents how we collect the historical issues of FNs and FPs from the issue repositories of the studied analyzers, and Section~\ref{sec:issue_analysis} explains how we manually analyze these issues.

\begin{figure}[t]
    \centering
    \includegraphics[height=4.6cm]{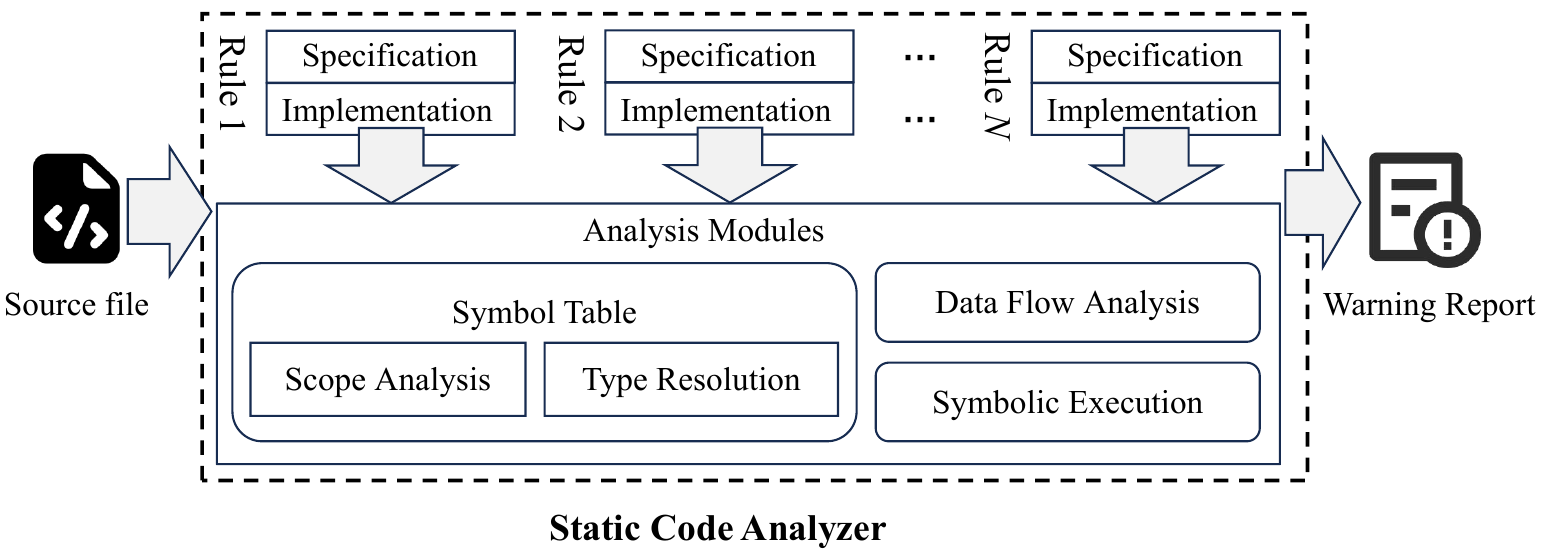}
    \vspace*{-1pc}
    \caption{Typical workflow and architecture of a static code analyzer.}
    \label{fig:toolFlowchart}
\end{figure}

\subsection{\rC{Background on Static Code Analyzers}}
\label{sec:background}
Figure \ref{fig:toolFlowchart} shows the typical workflow and architecture of classic static code analyzers like \textsc{PMD}~\cite{pmd}, \textsc{SpotBugs}~\cite{spotbugs}, and  \textsc{SonarQube}~\cite{SonarQube}.
Given the source files under check, an analyzer converts it to some form of intermediate representation, \eg, abstract syntax tree (AST), bytecode or control-/data-flow graph (CFG/DFG), and searches for the code snippets violating the rules (outputted as a warning report).
Specifically, an analyzer usually implements a number of rules, each of which has its own specification and implementation (\eg, \textsc{PMD}'s rules~\cite{pmd_java_rules}, \textsc{SpotBugs}'s rules~\cite{spotbugs_bug_descriptions}, \textsc{SonarQube}'s rules~\cite{sonarsource_java_rules}). Each rule is designed to detect one specific type of program flaws, \eg, \emph{best practice violations}, \emph{code design issues}, \emph{common programming mistakes} and \emph{security vulnerabilities}.
In detail, each rule is implemented based on some form of static analysis (chosen by the developers of the analyzers), \eg, AST-based syntactic pattern matching, data-flow analysis and symbolic execution.
\rB{Note that the major components of different analyzers may not align with each other.
    For example, all the classic analyzers like \textsc{PMD}, \textsc{SpotBugs} and \textsc{SonarQube} implement a number of checking rules (associated with the corresponding rule specifications),
    and all the analyzers would construct the symbol tables in their analysis modules.
    But only \textsc{SonarQube} uses symbolic execution for static analysis. \textsc{PMD} only implements reaching definition analysis (one specific data-flow analysis), while \textsc{SpotBugs} implements data-flow analysis more generally supporting forward and backward analyses. In addition, \textsc{PMD} and \textsc{SonarQube}'s analysis module works at the source code level, while \textsc{SpotBugs}'s works at the bytecode level.}

For example, \href{https://docs.pmd-code.org/latest/pmd\_rules\_java\_bestpractices.html\#abstractclasswithoutabstractmethod}{\textit{AbstractClassWithoutAbstractMethod}}~\cite{pmd_abstract_class_without_abstract_method} is one of the rules implemented in \textsc{PMD} to enforce generally accepted best practices. It warns any abstract class which does not contain any abstract methods because an abstract class suggests an incomplete implementation, which is to be completed by subclasses implementing the abstract methods. This rule is implemented by syntactic pattern matching based on AST.
For another example,
\href{https://rules.sonarsource.com/java/tag/symbolic-execution/RSPEC-2259/}{\textit{S2259 (Null pointers should not be dereferenced)}}~\cite{sonar_java_rspec_2259} is one of the rules in \textsc{SonarQube} to detect programming bugs. It checks that a reference to {\small\texttt{null}} should never be dereferenced or accessed because doing so will cause a {\small\texttt{NullPointerException}}.
This rule is implemented upon \textsc{SonarQube}'s symbolic execution engine.

\subsection{Static Code Analyzers Selected for Our Study}
\label{sec:studied_static_analyzers}
\rB{In this work, we focus on the static code analyzers for Java language because Java is one of the most popular programming language and targeted by (many) existing analyzers.}
Specifically, we selected the three representative static code analyzers for Java, \ie, \textsc{PMD}~\cite{pmd}, \textsc{SpotBugs}~\cite{spotbugs}, and  \textsc{SonarQube}~\cite{SonarQube}, as the subjects for our study based on the following reasons.
First, these tools are popular.
For example, \textsc{PMD} has been integrated into several industrial IDEs (\eg, Eclipse, IntelliJ IDEA, Visual Studio Code);
\textsc{SpotBugs}, the successor of \textsc{FindBugs}~\cite{findbugs}, has been used by Google;
\textsc{SonarQube} has been applied at the CI pipelines by many companies.
Moreover, these tools have been widely studied by many prior work~\cite{wang2022find,AreSonarQubeRules,howmanyof,usage,effectiveness_2,effectiveness_3,effectiveness_4}.
Second, these tools are open-sourced. It eases the issue collection and analysis. We can inspect the fixing patches and tool implementations to investigate FNs and FPs.
Third, these tools adopt different analysis strategies, including
syntactic pattern matching, data-flow analysis and symbolic execution.
These characteristics can help us gain more overall understanding on FNs and FPs.
\rM{There are other static code analyzers for Java~\cite{NachtigallSB22} like \textsc{FindBugs}~\cite{findbugs}, \textsc{ErrorProne}~\cite{errorprone}, \textsc{Infer}~\cite{infer} and \textsc{CheckStyle}~\cite{checkstyple}, but we did not select these tools as the subjects.
We did not select \textsc{FindBugs} because it is deprecated and has not been maintained for almost ten years (its latest version was released in 2015). \textsc{FindBugs}'s development has moved to \textsc{SpotBugs} (which we studied in this work). Moreover,  \textsc{FindBugs} and \textsc{ErrorProne} do not maintain the issues of FNs and FPs (\eg, without labeling the FNs/FPs or linking with the fixing patches), \textsc{Infer} has only a few checking rules (i.e., 25 rules) and thus few issues of FNs/FPs, and \textsc{Checkstyle} focuses more on coding standards rather than program flaws.}

\begin{figure}[t]
    \includegraphics[height=3cm]{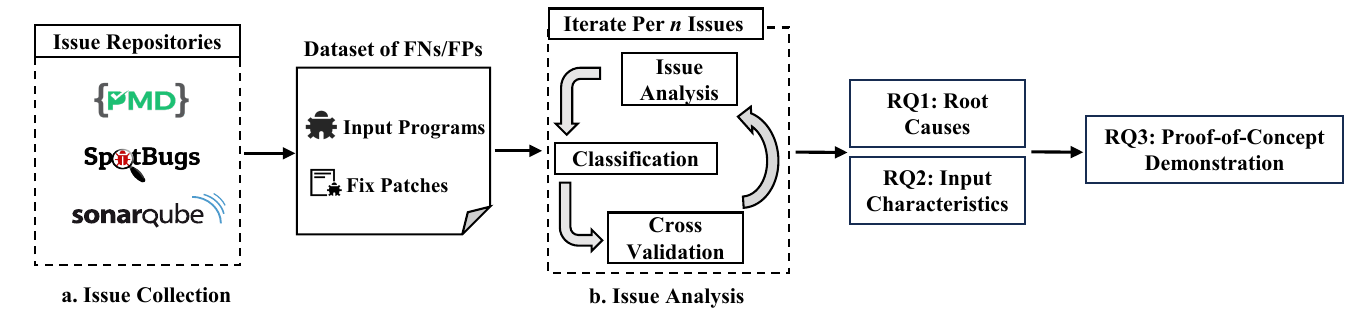}
    \vspace*{-1pc}
    \caption{Workflow of our study.}
    \label{fig:overview}
\end{figure}

\vspace{3pt}
Figure \ref{fig:overview} shows our study's workflow including collecting and analyzing the historical issues of FNs and FPs from the issue repositories of the studied analyzers. We explain the process in Section~\ref{sec:issue_collection} and Section~\ref{sec:issue_analysis}.

\subsection{Collecting Historical Issues of FNs and FPs}
\label{sec:issue_collection}

To obtain a broad range of issues, we collected \emph{all} the reported issues before the time of our study (which was started in Oct. 2022).
We focus on the FNs/FPs that were reported on checking Java programs supported by all the three studied analyzers. \rB{Moreover, we require that these FNs/FPs have been confirmed \emph{and} fixed by the developers because the confirmed bug information, the committed patches, the discussions between the developers and the implementations of the rules provide us informative details to study the root causes and the input characteristics of these issues.}

We detail the issue collection process below. \rB{Note that we focus on those issues of FNs and FPs which have been confirmed \emph{and} fixed.
    For \textsc{PMD}, an issue is considered as \emph{confirmed} if \textsc{PMD}'s developers explicitly labeled the issue with ``a:false-positive'' or ``a:false-negative''. For \textsc{SpotBugs} and \textsc{SonarQube}, an issue is considered as \emph{confirmed} if the developers explicitly confirmed the issue is a FN or a FP during discussions.
    An issue is considered as \emph{fixed} if the issue report has been closed and associated with the fixing commits.}

\begin{enumerate}[label=\textbullet,leftmargin=*]

    \item \textbf{\textsc{PMD}}.
          We collected all the \emph{closed} issues with the labels \textit{``a:false-positive''} and \textit{``a:false-negative''} from \textsc{PMD}'s issue repository on GitHub. To collect issues that are related to Java programs, we filtered out the issues that do not contain the keyword \textit{``java''} in the issue title or body. We initially obtained 80 FNs and 228 FPs.

    \item \textbf{\textsc{SpotBugs}}.
          \textsc{SpotBugs}'s issues are not well-classified with explicit issue labels.
          Thus, we checked whether the issue title or body contains the keywords \textit{``FP''}, \textit{``FN''}, \textit{``bug''}, \textit{``error''}, \textit{``false positive''}, \rB{\textit{``false alarm''}}, or \textit{``false negative''} to filter issues. Finally, we obtained 104 unclassified issues.

    \item \textbf{\textsc{SonarQube}}.
          \textsc{SonarQube}'s issues are maintained on the community website.
          We use the APIs of \textsc{Sonar} community~\cite{sonarCommunity} to collect the issues with the labels \textit{``Clean Code''}, \textit{``Report False-positive / False-negative''}, \textit{``java''}, \textit{``answered''}, and \textit{``closed''}.
          We obtained 229 unclassified issues.

\end{enumerate}

\begin{table}[t]
    \caption{Statistics of the studied static code analyzers (K=1,000), and the dataset of valid FNs and FPs.}
    \vspace*{-1pc}
    \label{tab:issueCollection}
    \centering
    \footnotesize
    \begin{adjustbox}{scale=1.0}
        \begin{tabular}{>{\centering\arraybackslash}m{1.5cm}>{\centering\arraybackslash}m{1cm}>{\centering\arraybackslash}m{1.5cm}>{\centering\arraybackslash}m{2cm}>{\centering\arraybackslash}m{2cm}>{\centering\arraybackslash}m{2cm}}
            \toprule
            \textbf{Analyzer}           & \textbf{GitHub \#Stars} & \textbf{\#Total Java Rules} & \textbf{Date of first collected issue} & \textbf{Date of last collected issue} & \textbf{\#Valid Issues (FP/FN)} \\
            \midrule
            \textit{\textsc{PMD}}       & 4.3K                    & 325                         & 2017/2/17                              & 2022/10/8                             & 226(155/71)                     \\
            \textit{\textsc{SpotBugs}}  & 3K                      & 468                         & 2016/12/7                              & 2022/10/27                            & 21(15/6)                        \\
            \textit{\textsc{SonarQube}} & 7.7K                    & 618                         & 2018/6/11                              & 2022/10/30                            & 103(100/3)                      \\
            \bottomrule
        \end{tabular}
    \end{adjustbox}

\end{table}

We manually verified whether each collected issue is a valid FN or FP by inspecting the issue report and reproducing the issue if necessary.
During this process, we excluded those mislabeled or duplicated issues,
won't fix issues of deprecated rules, and the issues caused by user misconfigurations.
We only retained those developer-confirmed \emph{and} -fixed FNs/FPs.
Finally, we obtained a dataset of 350 FNs/FPs, including 71 FNs/155 FPs of \textsc{PMD}, 6 FNs/15 FPs of \textsc{SpotBugs}, and 3 FNs/100 FPs of \textsc{SonarQube}.
Each issue is associated with at least one piece of issue-triggering program from the issue report.
Table~\ref{tab:issueCollection} gives the detailed statistics of the collected issues.
Note that many issues of \textsc{Spotbugs}
were excluded because they were not confirmed by the developers, \eg,
the issue reporters misunderstood or misconfigured the rules, or the issue-triggering programs were deemed as unrealistic (not likely written by humans).
\rB{The issues which were closed but not confirmed were also excluded from our study.}

\subsection{Analyzing Historical Issues of FNs and FPs}
\label{sec:issue_analysis}

To answer RQ1 and RQ2, we investigated 350 FNs/FPs in the dataset to understand the root causes and input characteristics.
Specifically, we inspected the documentation (\eg, rule specifications), the discussions in the issue reports, the issue-triggering program and the fixing patches.
In this process, some issue labels (\eg, \emph{dataflow analysis}) or some keywords (\eg, \emph{anonymous class}) in the issue titles could indicate the likely root causes or the input characteristics leading to the issue.
To confirm our understanding, we inspected the fixing patches (\ie, fixing commits or pull requests) and reproduced the issue against the rule implementation if necessary. During the analysis process, we may inspect the AST of the issue-triggering program and mutate the program to locate which code parts are pivot for triggering FNs/FPs.

Specifically, to build the taxonomies, we adopted the open card sorting approach~\cite{spencer2004card} and conducted the preceding analysis in an iterative process. In each iteration, 30 issues in the dataset were randomly selected and two of the co-authors independently studied each of these issues.
These two co-authors are familiar with Java (with 5 years of Java programming experience) and the relevant static analysis techniques.
According to their own understanding, they independently labeled
each issue with the categories of root causes and input characteristics.
Afterward, these two co-authors cross-validated and discussed
the labels until they reached a consensus on the categorized results.
When they could not reach a consensus, the other co-authors
participated in the discussion to help make the final decision. Such
an iteration was repeated twelve times until all 350 issues were
analyzed. We observed that this iterative process converged on
the categories after the first 3 rounds.
This manual analysis process requires considerable knowledge of Java and the implementations of static code analyzers, which took us around six person months.
\rC{During this manual analysis process, we computed the Cohen's Kappa coefficient~\cite{kappa} to evaluate the inter-rater agreement~\cite{inter-rater} between the two co-authors.
Cohen's Kappa is a statistical measure used to quantify the level of agreement between two raters beyond what would be expected by chance. It is particularly useful when evaluating subjective judgments, where the goal is to determine how consistently two or more raters classify or label the same items.
The high inter-rater agreement indicates that the raters are in close alignment in their judgments, while low agreement suggests differences in their evaluations. By applying Cohen's Kappa, we ensured that the categories of root causes and input characteristics assigned by the co-authors were consistent and reliable, minimizing the impact of individual biases.
In the first three rounds, the Cohen's Kappa coefficient was around 0.5, primarily because these two co-authors were still unfamiliar with the FNs and FPs of the static analyzers and hadn't yet reached a consensus. As their understanding deepened and they gained more experience, the coefficient increased to 0.9 in subsequent rounds. After the discussions between them in each round, they eventually achieved a perfect agreement with a Kappa of 1.0.}

\rB{
    Our study primarily focuses on the root causes and input characteristics of FNs and FPs, and has not explored other possible dimensions (\eg, severity, affected versions, latency, and fixability) because these other dimensions are difficult to be analyzed or may bring additional threats to validity. For example, none of \textsc{PMD}, \textsc{SpotBugs} and  \textsc{SonarQube} maintains the labels of severity because how to tell the severity of a FN or a FP is unclear.
    Investigating the versions of the analyzers affected by the issues of FNs and FPs may not offer interesting insights because the FNs/FPs were reported for the checking rules. Instead, examining how many times of the rules were affected by FNs/FPs might be more interesting (which we have investigated in Section~\ref{sec:analyzer_weaknesses}).
    The latency of fixing FNs/FPs may or may not indicate the difficulty of fixing
    FNs/FPs. The fixability is also difficult to measure in an objective manner because some issues were fixed by workarounds.
}

\begin{table*}[t]
    \footnotesize
    \caption{The taxonomy of root causes of FNs and FPs from the three static code analyzers.}

    \vspace*{-1pc}
    \label{tab:rootcause_categories}
    \centering
    \begin{tabular}{Y{4.2cm}|Y{5cm}|c|c|c}
        \rowcolor{headerColor}
        \textbf{Category}                                                                                                   & \textbf{Description}                                                                             & \textbf{Example}           & \textbf{\#Issues} & \textbf{Ratio}                                                                                                                                                                                     \\ \hline
        \textcolor{blue}{\emptycirc} \textcolor{darkgreen}{\emptycirc} \textbf{Flawed Rule Specification}                   & Some design flaws in the rule specification.                                                     & -                          & 21                & 6\%                                                                                                                                                                                                \\
        \rowcolor{gray!20}
        \textcolor{blue}{\halfcirc} \textcolor{darkgreen}{\emptycirc} \textbf{Inconsistent Rule Implementation}             & The rule specification is correct but the implementation is inconsistent with the specification. & -                          & 9                 & 2.6\%                                                                                                                                                                                              \\
        \hspace{4ex}  \phantom{  }\textbf{Unhandled Language Features or Libraries}                                         & Unhandled Language Features or Libraries                                                         & -                          & 101               & 28.9\%                                                                                                                                                                                             \\
        \textcolor{blue}{\emptycirc} \textcolor{darkgreen}{\halfcirc} \quad Unhandled Language Features                     & Some language features (\eg lambda expressions) are not handled.                                 & Figure~\ref{fig:examples}a & 74                & 21.1\%                                                                                                                                                                                             \\
        \textcolor{blue}{\emptycirc} \textcolor{darkgreen}{\emptycirc} \quad Unhandled Java Libraries                       & Some Java libraries (\eg java.util.Optional) are not handled.                                    & -                          & 27                & 7.7\%                                                                                                                                                                                              \\
        \rowcolor{gray!20}
        \hspace{4ex}  \phantom{  }\textbf{Missing Cases}                                                                    & Missing Cases                                                                                    & -                          & 125               & 35.7\%                                                                                                                                                                                             \\
        \rowcolor{gray!20}
        \textcolor{blue}{\emptycirc} \textcolor{darkgreen}{\emptycirc} \quad Missing cases that should be whitelisted       & Some objects that should be added to the whitelist are missing.                                  & Figure~\ref{fig:examples}b & 38                & 10.9\%                                                                                                                                                                                             \\
        \rowcolor{gray!20}
        \textcolor{blue}{\halfcirc} \textcolor{darkgreen}{\emptycirc} \quad Missing cases which should be similarly handled & Missing cases or objects which should be similarly handled.                                      & -                          & 30                & 8.6\%                                                                                                                                                                                              \\
        \rowcolor{gray!20}
        \textcolor{blue}{\halfcirc} \textcolor{darkgreen}{\emptycirc} \quad Missing specific cases                          & Missing specific cases that are not covered by the two categories above                          & -                          & 57                & 16.3\%                                                                                                                                                                                             \\
        \textcolor{blue}{\emptycirc} \textcolor{darkgreen}{\halfcirc} \textbf{Mishandling Intermediate Representations}     & Matching the incorrect nodes during the traversal of the IR (\eg AST)                            & -                          & 24                & 6.9\%                                                                                                                                                                                              \\
        \rowcolor{gray!20}
        \hspace{4ex}  \phantom{  } \textbf{Analysis Module Error or Limitation}                                             & Analysis Module Error or Limitation                                                              & -                          & 51                & 14.6\%                                                                                                                                                                                             \\
        \rowcolor{gray!20}
        \textcolor{blue}{\emptycirc} \textcolor{darkgreen}{\emptycirc} \quad Scope analysis error or limitation             & Wrong resolving of the variable scopes.                                                          & Figure~\ref{fig:examples}c & 6                 & 1.7\%                                                                                                                                                                                              \\
        \rowcolor{gray!20}
        \textcolor{blue}{\emptycirc} \textcolor{darkgreen}{\emptycirc} \quad Type resolution error or limitation            & Incorrect or limited type resolution.                                                            & Figure~\ref{fig:examples}d & 38                & 10.9\%                                                                                                                                                                                             \\
        \rowcolor{gray!20}
        \textcolor{blue}{\emptycirc} \textcolor{darkgreen}{\emptycirc} \quad Dataflow analysis error or limitation          & Errors in the dataflow analysis or limited use of dataflow analysis                              & Figure~\ref{fig:examples}e & 3                 & 0.9\%                                                                                                                                                                                              \\
        \rowcolor{gray!20}
        \textcolor{blue}{\emptycirc} \textcolor{darkgreen}{\emptycirc} \quad Symbolic execution error or limitation         & Errors or limitations in the symbolic execution engine.                                          & Figure~\ref{fig:examples}f & 4                 & 1.1\%                                                                                                                                                                                              \\
        \textcolor{blue}{\emptycirc} \textcolor{darkgreen}{\emptycirc} \textbf{General Programming Error}                   & General errors which are not unique for static code analyzers.                                   & -                          & 11                & 3.1\%                                                                                                                                                                                              \\
        \rowcolor{gray!20}
        \hspace{4ex}  \phantom{  } \textbf{Miscellaneous}                                                                   & Minor issues that affect only a few FNs/FPs.                                                     & -                          & 8                 & 2.3\%                                                                                                                                                                                              \\ \hline
        \multicolumn{5}{p{13.5cm}}{The symbols``\textcolor{blue}{\halfcirc}'' and  ``\textcolor{blue}{\emptycirc}'', ``\textcolor{darkgreen}{\halfcirc}'' and ``\textcolor{darkgreen}{\emptycirc}''  denote the differences between our study and the two most relevant work from Wang \etal ~\cite{wang2022find} and Zhang \etal~\cite{Statfier}, respectively, in terms of the results of issue analysis. We discuss the differences in detail in Section~\ref{sec:related_work}.} \\
    \end{tabular}

\end{table*}

%% file: Sections/RQ1.tex
\section{RQ1: Root Causes}
\label{sec:rootcause}
In this section, we focus on the 350 issues in the three analyzers to study the root causes.
We identified the 7 major root causes from 350 FNs/FPs. To be concise, we brief these root causes in Table \ref{tab:rootcause_categories}.
Specifically, we categorized the issues into the disjoint groups of root causes according to the major components of a static code analyzer (see Figure~\ref{fig:toolFlowchart}) in which the issues may happen.
We explain and illustrate these root causes as follows.

\subsection{Flawed Rule Specification}
\label{rtcs:spec}

In this category, the rule specification itself (documented in natural language) is flawed, thus leading to FNs or FPs.
For example, \textsc{PMD}'s rule \textit{DoNotUseThreads}~\cite{pmd_donotusethreads} intends to warn against the direct use of threads (\eg, {\small\texttt{Thread}}, {\small\texttt{ExecutorService}}) in favor of J2EE's managed thread mechanism. However, this rule is incorrectly designed to warn the use of {\small\texttt{Runnable}} because the developers misunderstand that {\small\texttt{Runnable}} is identical to {\small\texttt{Thread}}.
In fact, {\small\texttt{Runnable}} is a class whose instances are intended to be executed by a thread.
It is compliant with the managed thread environment in J2EE, and thus should not be warned.
Due to this flawed specification, this rule reports a FP (\textsc{PMD}'s Issue \#1627) when the input program uses {\small\texttt{Runnable}}.
To resolve this issue, the developer fixed the rule specification and implementation\footnote{https://github.com/pmd/pmd/pull/2078/commits/5739041b164d0bb4cc94715aa3bed801c4565e02}.

\begin{tcolorbox}[
        colframe=black,
        boxrule=0.5pt,top=0.1mm,bottom=0.1mm,left=1mm
    ]
    \textbf{Finding 1 \& Implications}: \textit{Flawed Rule Specification} could lead to both FNs and FPs. To avoid such rule specification issues, the analyzers' developers should carefully inspect and fully understand the language specifications (\eg, Java language) \rA{as well as the language's idioms and the best practices} when designing the rules.
\end{tcolorbox}

\begin{figure*}[t]
    \includegraphics[width=1\textwidth]{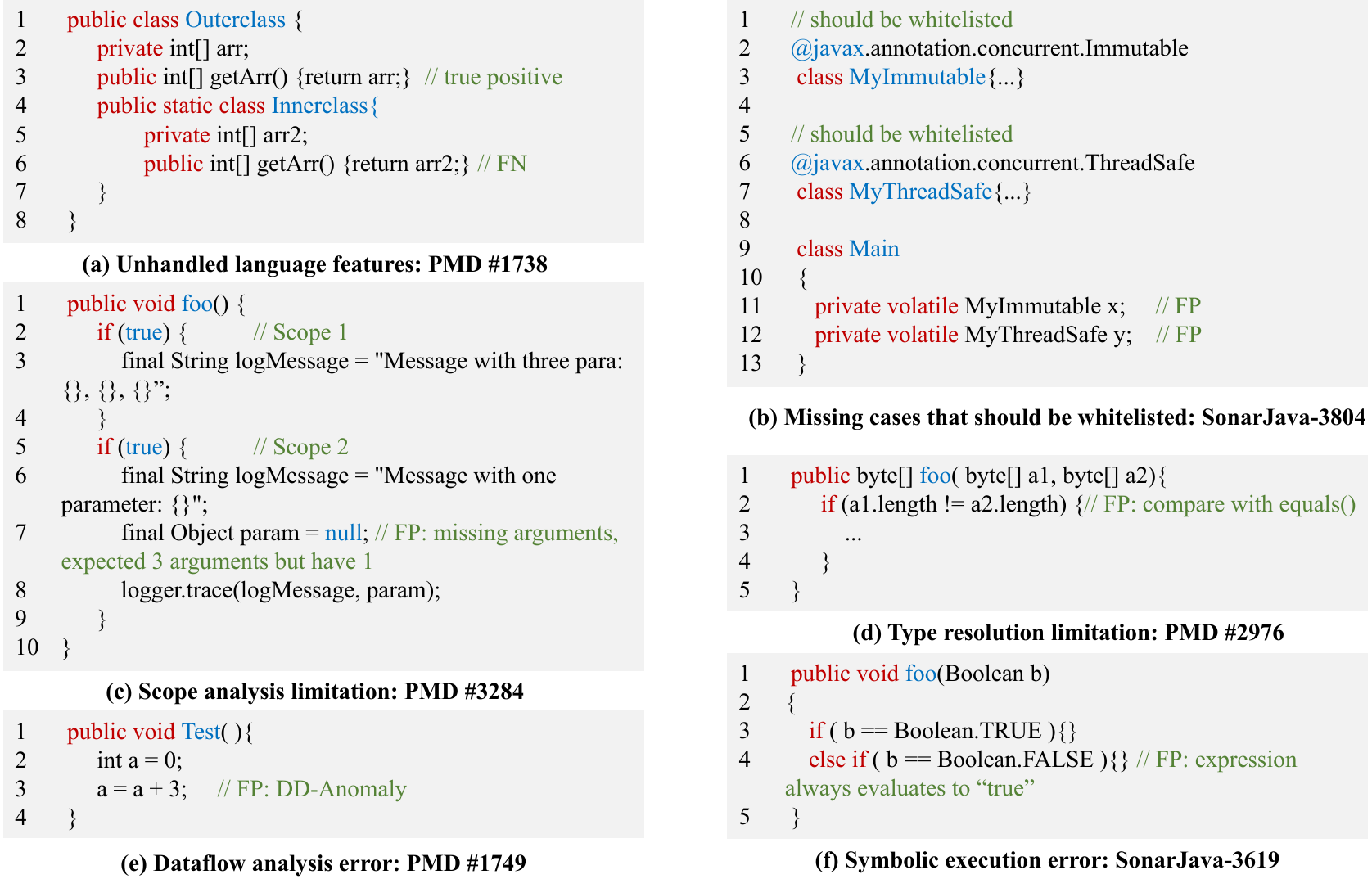}
    \caption{Illustrative examples for explaining root causes (the code snippets are simplified).}
    \label{fig:examples}
\end{figure*}

\subsection{Inconsistent Rule Implementation}
\label{rtcs:inconsistencies}
In this category, the rule specification is correct but
the implementation is \emph{inconsistent}.
For example, \textsc{PMD}'s rule \textit{AvoidThrowingNullPointerException} is specified to warn manually throwing {\small\texttt{NullPointerException}}s.
However, this rule's implementation is inconsistent with the specification. It simply warns every occurrence of {\small\texttt{NullPointerException}} without checking whether an exception throwing occurs. As a result, this rule leads to a FP (\textsc{PMD}'s Issue \#2580): for example, the rule will warn this code line {\small\texttt{Exception e = new NullPointerException("Test message")}}, however this code line
does not throw an exception but only creates an exception object.

\subsection{Unhandled Language Features or Libraries}
\label{rtcs:unhandled lfol}

\subsubsection{Unhandled Language Features}
\label{sec:unhandled_language_features}
A rule may lead to FNs or FPs if some language features (\eg, lambda expressions, nested classes) are not handled.
For example, Figure~\ref{fig:examples}a showcases a FN of \textsc{PMD}'s rule \textit{MethodReturnsInternalArray} induced by nested classes.
This rule warns the methods that return internal arrays. In this case, {\small\texttt{arr}} at line 2 and {\small\texttt{arr2}} at line 5 are internal arrays. The two methods {\small\texttt{getArr()}}s at line 3 and 6, return the internal arrays {\small\texttt{arr}} and {\small\texttt{arr2}}, respectively. They violate the rule. However, the rule did not handle the nested class, and did not warn at line 6.

\subsubsection{Unhandled Java Libraries}
\label{rtcs:unhandled_java_libraries}
Some Java libraries were not handled, leading to FNs or FPs. For example, Java 8 introduces the class {\small\texttt{Optional}} to handle optional values.
Specifically, {\small\texttt{Optional}} provides {\small\texttt{Optional.isPresent()}} to check if a value is present (\ie, {\small\texttt{non-null}}). Later, Java 11 introduced a new method {\small\texttt{Optional.isEmpty()}}. This method allows to check whether the {\small\texttt{Optional}} value is {\small\texttt{null}}.
However, \textsc{SonarQube} failed to timely support {\small\texttt{Optional.isEmpty()}}, leading to a FP (see SonarQube's issue SonarJava-3087).

\begin{tcolorbox}[
        colframe=black,
        boxrule=0.5pt,top=0.2mm,bottom=0.2mm,left=1mm
    ]
    \textbf{Finding 2 \& Implications}: \textit{Unhandled language features or libraries} affects 101 of the 350 issues (28.9\%), which is one major root cause. It indicates that the analyzers' developers should \emph{timely} check the rules when some new Java language features or new Java libraries are introduced, and update the rule implementations if necessary.
\end{tcolorbox}

\subsection{Missing Cases}
\label{rtcs:mc}
\subsubsection{Missing cases that should be whitelisted}
\label{rtcs:whitelist}
The analyzers commonly use whitelists (precluding checking on specific code elements) to avoid spurious warnings.
However, these analyzers may miss the cases which should be whitelisted.
For example, Figure~\ref{fig:examples}b showcases a FP of \textsc{SonarQube}'s rule \textit{S3077} due to failing to whitelist some Java annotations.
This rule warns the non-primitive fields modified by {\small\texttt{volatile}}.
In this example, the classes {\small\texttt{MyImmutable}} and {\small\texttt{MyThreadSafe}} are annotated as immutable (line 2) and thread-safe (line 6), respectively. It means the immutability and the thread-safety are already ensured by the users.
Thus, the non-primitive fields (\ie, {\small\texttt{x}} at line 11 and {\small\texttt{y}} at line 12) should not be warned. The developers fixed this FP by adding the annotations {\small\texttt{@ThreadSafe}} and {\small\texttt{@Immutable}} into the whitelist of this rule.
\emph{Note that the issues of this category all lead to FPs}.

\subsubsection{Missing cases which should be similarly handled}
\label{rtcs:incomplete enumeration}

Some cases are functionally equivalent \wrt the rule specification.
However, sometimes analyzers may only handle some cases but fail to similarly handle
others. For example, \textsc{SonarQube}'s issue SonarJava-3586 reported a FP because \textsc{SonarQube} only handles the annotation {\small\texttt{org.springframework.lang.Nullable}} but not the annotation {\small\texttt{reactor.util.annotation.Nullable}}. Both of these annotations have identical semantics, \ie, indicating the annotated elements can be {\small\texttt{null}} in some circumstances.
To fix this issue, the developers added {\small\texttt{reactor.util.annotation.Nullable}}.
\rB{
    Note that this issue does not belong to the category of unhandled language features (discussed in Section~\ref{sec:unhandled_language_features}) because the rule can correctly handle the @Nullable annotations but missed {\small\texttt{reactor.util.annotation.Nullable}}.
}

\subsubsection{Missing specific cases}
\label{rtcs:missingcases}
The analyzers may miss other specific cases when doing rule checking.
For example, \href{https://github.com/pmd/pmd/issues/2275}{\textsc{PMD}'s Issue \#2275} showcases a FP of the rule \textit{AppendCharacterWithChar}.
For an object of class {\small\texttt{StringBuffer}}, say {\small\texttt{sb}}, this rule recommends converting {\small\texttt{sb.append(``a'')}} to {\small\texttt{sb.append(`a')}} to improve performance. However, for the case of {\small\texttt{sb.append(``a''.repeat(length))}}, it is valid and should not be warned by this rule. But \textsc{PMD} assumes that {\small\texttt{append()}}'s argument should always be string literals. It does not consider the case of using method calls like {\small\texttt{repeat()}}.

\begin{tcolorbox}[
        colframe=black,
        boxrule=0.5pt,top=0.2mm,bottom=0.2mm,left=1mm
    ]
    \textbf{Finding 3 \& Implications}: \textit{Missing cases} affects 125 of the 350 issues (35.7\%), which is the most common root cause. It indicates that the analyzers' developers should (1) carefully decide which cases need to be whitelisted and similarly handled, and (2) design different diverse test programs to validate the rule implementations.
\end{tcolorbox}

\subsection{Mishandling Intermediate Representations}
\label{rtcs:ir}
Static code analyzers usually convert input programs into some
intermediate representation (IR), \eg, abstract syntax trees or bytecode, for analysis. However, mishandling IRs
could lead to FNs or FPs.
For example, \textsc{PMD}'s Issue \#3949 reports a FN of rule \textit{FinalFieldCouldBeStatic}. The rule warns if a {\small\texttt{final}} field is assigned by a compile-time constant but not modified by {\small\texttt{static}}.
This rule correctly warns {\small\texttt{public final int BAR = 42}}, but fails to warn {\small\texttt{public final int BAR = (42)}}. Because  the IR structures of these two cases are different at the level of abstract syntax tree due to the parenthesis.

\subsection{Analysis Module Error or Limitation}
\label{rtcs:analysis module}

\subsubsection{Scope analysis error or limitation}
\label{subsection:scope_analysis}
Java adopts the static scoping rules to analyze the scopes of symbols.
However, some issues are caused by imprecise scope analysis. Figure~\ref{fig:examples}c showcases a FP of \textsc{PMD}'s rule \textit{InvalidLogMessageFormat}. The rule checks whether the numbers of arguments and placeholders ( ``{\small\texttt{\{\}}}'') in \textsc{slf4j} or \textsc{log4j2} loggers are matched. In this case, the variable {\small\texttt{logMessage}} in Scope 1 (lines 2 to 4) is incorrectly mapped to the variable {\small\texttt{logMessage}} in Scope 2 (lines 5 to 9).
As a result, the rule incorrectly warns that the argument {\small\texttt{param}} at line 7 is not matched with the three placeholders in the {\small\texttt{logMessage}} (in Scope 1).

\label{rtcs:icr}

\subsubsection{Type resolution error or limitation}
\label{subsection:type_resolution_err}

Type resolution is a crucial ability of analyzers to decide the types of symbols. However, the implementation of type resolution may have some limitations.
Figure~\ref{fig:examples}d showcases a FP of \textsc{PMD}'s rule \textit{CompareObjectsWithEquals} due to the incorrect type resolution. This rule requires using ``{\small\texttt{equals()}}'' instead of ``{\small\texttt{==}}'' to compare object references. However, at line 2, {\small\texttt{a1.length}} and {\small\texttt{a2.length}} are the type of integers and it is fine to use ``{\small\texttt{==}}''. \textsc{PMD}'s incorrect type resolution leads to a FP.
\rA{We observe that these three analyzers do not utilize Java compilers for precise type resolution, but implement their own type resolutions.
}

\begin{figure}[t]
    \centerline{\includegraphics[width=1\textwidth]{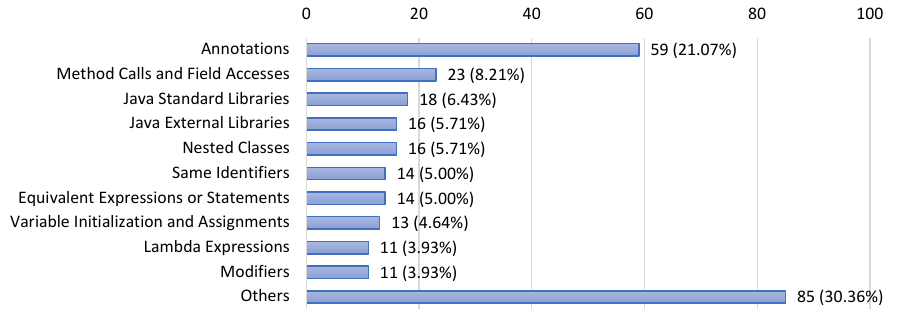}}
    \vspace{-1pc}
    \caption{Input characteristics leading to FNs/FPs.}
    \label{fig:proportion}
\end{figure}
\subsubsection{Dataflow analysis error or limitation}
Some dataflow analysis errors
could lead to FNs or FPs.
For example, Figure~\ref{fig:examples}e showcases a FP of \textsc{PMD}'s rule \textit{DataflowAnomalyAnalysis}. This rule warns about data flow anomalies, \eg, redefinition of a recently defined variable without prior usage. In this example, the variable {\small\texttt{a}} is defined at line 2, used at line 3, and then redefined at line 3.
No data flow anomalies happen. However, the dataflow analysis module erroneously processes the variable definition and use from left to right at line 3 (assuming {\small\texttt{a}} is defined and then used).

\subsubsection{Symbolic execution error or limitation}
\textsc{SonarQube} uses symbolic execution to implement some rules.
Some FNs and FPs are caused by the errors in the symbolic execution engine. For example, Figure~\ref{fig:examples}f showcases a FP of \textsc{SonarQube}'s rule \textit{S2589}. This rule warns the gratuitous boolean expressions that always evaluate to ``true'' or ``false''. In this example, the symbolic execution engine misinterprets that the boxed type {\small\texttt{Boolean}} has only two kinds of values {\small\texttt{Boolean.TRUE}} and {\small\texttt{Boolean.FALSE}} but missed the other value {\small\texttt{null}}. When the engine meets the condition {\small\texttt{if(b == Boolean.TRUE)}} at line 3, it tries to reach the {\small\texttt{else if}} branch with the ``only'' other value {\small\texttt{Boolean.FALSE}}, and thus incorrectly assumes that the boolean expression at line 4 always evaluates to ``true''.

\begin{tcolorbox}[
        colframe=black,
        boxrule=0.5pt,top=0.2mm,bottom=0.2mm,left=1mm
    ]
    \textbf{Finding 4 \& Implication}: \textit{Type resolution errors or limitations} is the most common root cause among the static analysis modules. It indicates that improving the abilities and precision of such static analysis is important to mitigate FNs/FPs.
\end{tcolorbox}

\subsection{General Programming Error}
\label{rtcs:generalpe}
Some FNs or FPs are caused by general programming errors, \eg, mistaking the logic expression {\small\texttt{(A || B)}} as {\small\texttt{(A \&\& B)}}, where {\small\texttt{A}} and {\small\texttt{B}} are boolean conditions.

\subsection{Miscellaneous}

This category includes some miscellaneous reasons.
For example, some issues are caused by the unhandled whitespace in rule properties, some are caused by missing setting the property value of target type, and some are the limitations of libraries used by the analyzers (\eg limitations of \textsc{XPath version 1.0} used by \textsc{PMD}).
\rB{These cases are not relevant to the core functionality of the analyzers in performing rule checking.}

%% file: Sections/RQ2.tex
\section{RQ2: Input Characteristics}
\label{sec:characteristic}

This section investigates the input characteristics leading to FNs and FPs of the studied analyzers.
From the 350 issues, we excluded 70 issues because these issues are not relevant to the input programs (\eg, those issues caused by \emph{flawed rule specification}, \textit{inconsistent rule implementation}, \textit{general programming errors}). Thus, we analyzed the remaining 280 issues and identified 10 major input characteristics. Figure~\ref{fig:proportion} shows their proportions. We illustrate these input characteristics from the most to the least common and also discuss their correlations with the root causes.

\begin{figure*}[t]

    \includegraphics[width=1\textwidth]{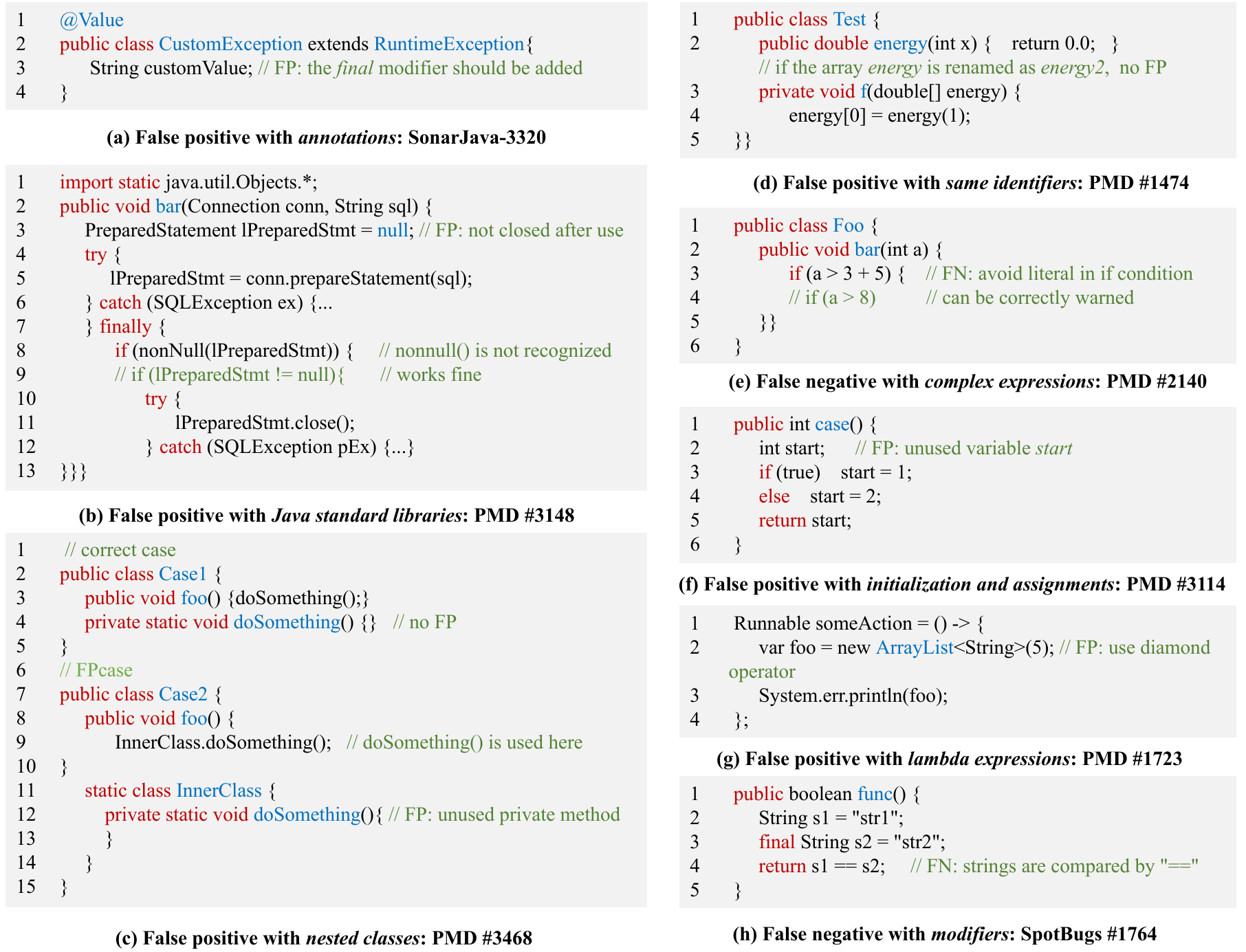}
    \centering
    \caption{Illustrative examples for explaining input characteristics (the code snippets are simplified)}
    \label{fig:characteristics}
\end{figure*}

\vspace{2pt}
\noindent\textbf{Annotations.}
Java annotations are a form of metadata that provides additional information of Java programs.
They are
usually placed above the declarations of code segments (\eg, classes, methods and fields) to provide compile-time or runtime information.
Figure~\ref{fig:characteristics}a showcases this characteristic which leads to a FP of \textsc{SonarQube}'s rule \textit{S1165}. This rule requires that the fields of exception classes should be {\small\texttt{final}}. When {\small\texttt{@Value}} (from {\small\texttt{Lombok}}) annotates the class {\small\texttt{CustomException}} (line 1), all the fields (\eg, {\small\texttt{customValue}} at line 3) in the class are made {\small\texttt{final}} by default. However, the rule does not handle {\small\texttt{@Value}}, and thus warns that a {\small\texttt{final}} modifier should be added (a FP).
\rB{Note that this category of ``annotations'' are mainly triggered by the root causes of \emph{missing cases that should be whitelisted} (Section~\ref{rtcs:whitelist}) and \emph{unhandled Java libraries} (Section~\ref{rtcs:unhandled_java_libraries}).}

\begin{tcolorbox}[
        colframe=black,
        boxrule=0.5pt,top=0.2mm,bottom=0.2mm,left=1mm
    ]
    \textbf{Finding 5 \& Implication}: \textit{Annotations} is the most common input characteristics of programs to trigger FNs/FPs. It is mainly related with the root cause of \emph{missing cases that should be whitelisted} and affects all the three studied analyzers. The analyzers' developers should carefully model these annotations in the rule implementations.
\end{tcolorbox}

\vspace{2pt}
\noindent\textbf{Method Calls and Field Accesses.}
Method calls or field accesses may induce FNs or FPs.
This characteristic is mainly related with the root cause of \emph{type resolution errors or limitations} (Section~\ref{subsection:type_resolution_err}).
For example, \textsc{PMD} rule \textit{UseEqualsToCompareStrings} warns comparing strings by using ``{\small\texttt{==}}'' or ``{\small\texttt{!=}}''. In \textsc{PMD} issue \#3004, this rule gives
a FP on the expression {\small\texttt{s.charAt(0) == s.charAt(1)}}, where {\small\texttt{s}} is a {\small\texttt{string}} variable
and {\small\texttt{s.charAt(i)}} is a method call returning the character at index {\small\texttt{i}} of {\small\texttt{s}}.
This FP is caused by the incorrect type resolution on the return value of {\small\texttt{charAt()}}.
Figure~\ref{fig:examples}e shows another example of FP with the characteristics of accessing the field {\small\texttt{length}} of an {\small\texttt{array}}.

\vspace{2pt}
\noindent\textbf{Java Standard Libraries.}
Static code analyzers may fail to handle the classes in Java standard library (\eg, {\small\texttt{java.lang.*}}).
This characteristic is mainly related with the root cause of \emph{unhandled java libraries} (Section~\ref{rtcs:unhandled_java_libraries}).
For example,
Figure~\ref{fig:characteristics}b shows a FP of \textsc{PMD}'s rule \textit{CloseResource} when handling {\small\texttt{java.util.Objects.nonNull()}}. This rule warns unclosed resources. In Figure~5b, the resource {\small\texttt{lPreparedStmt}} is created (line 5) and properly closed (line 11). However, \textsc{PMD} reports a FP as it fails to
handle the semantic of  {\small\texttt{nonNull()}}, thus believes {\small\texttt{lPreparedStmt.close()}} is not reachable. Replacing {\small\texttt{nonNull()}} with {\small\texttt{lPreparedStmt!=null}} eliminates the FP.

\vspace{2pt}
\noindent\textbf{Nested Classes.}
Some rules may fail to support or handle nested classes.
This characteristic is mainly related with the root cause of \emph{unhandled language features} (Section~\ref{sec:unhandled_language_features}).
Figure~5c illustrates a FP of \textsc{PMD}'s rule \textit{UnusedPrivateMethod}. This rule warns unused private methods. For the private method {\small\texttt{doSomething()}} in class {\small\texttt{Case1}}, PMD does not report a warning at line 4. However, when {\small\texttt{doSomething()}} is declared in a nested class {\small\texttt{InnerClass}} (line 12) but used in {\small\texttt{foo}} (line 9), a FP (line 12) occurs.

\vspace{2pt}
\noindent\textbf{Same Identifiers.}
In Java programs, variables, fields, methods may have the same symbol names.
This characteristic is mainly related with the root causes of \emph{type resolution errors or limitations} (Section~\ref{subsection:scope_analysis})
and \emph{type resolution errors or limitations} (Section~\ref{subsection:type_resolution_err}).
Because the same identifiers may complicate the symbol table construction which requires
scope analysis and type resolution.
Figure~\ref{fig:characteristics}d shows an input program that triggers a FP of \textsc{PMD}'s rule \textit{ArrayIsStoredDirectly}. 
\rA{This rule warns that the constructors and methods receiving arrays should clone objects and store the copy. Because it can prevent future changes from the users affecting the original array. However, in Figure~\ref{fig:characteristics}d, when the double type array {\small\texttt{energy}} (line 3) has the same symbol name as the method {\small\texttt{energy}} (line 2), this rule mistakes the method call {\small\texttt{energy}} (on the right hand side, line 4) as an array, and thus reports a spurious warning (FP).}

\vspace{2pt}
\noindent\textbf{Complex Expressions or Statements.}
Some complex expressions or statements may induce FNs or FPs.
This characteristic is mainly related with the root causes of \emph{mishandling intermediate representations} (Section~\ref{rtcs:ir})
and \emph{missing specific cases} (Section~\ref{rtcs:missingcases}).
For example, complex arithmetic operations (\eg changing {\small\texttt{a>3}} to {\small\texttt{a>1+2}}) or complex boolean operations (\eg changing {\small\texttt{false}} to {\small\texttt{false||false}}),
addition of no side-effect expressions (\eg, encapsulating statements by {\small\texttt{if(true)}} or expressions by {\small\texttt{\{\}}}), and addition of {\small\texttt{this.}} to non-static field access (\eg changing {\small\texttt{a}} to {\small\texttt{this.a}}).
Figure~\ref{fig:characteristics}e shows a FN due to the complex arithmetic operation. While the rule successfully detects simple expression {\small\texttt{a>8}}, it fails to accurately analyze complex but equivalent expression {\small\texttt{a>3+5}}.

\begin{tcolorbox}[
        colframe=black,
        boxrule=0.5pt,top=0.2mm,bottom=0.2mm,left=1mm
    ]
    \textbf{Finding 6 \& Implication}: Deliberately complicating expressions or statements could be a useful strategy to stress-testing of the rule implementations and manifest the issues caused by \emph{mishandling intermediate representations} and \emph{missing specific cases}.
\end{tcolorbox}

\vspace{2pt}
\noindent\textbf{Java External Libraries.}
Java external libraries are prepackaged modules in JAR files, offering versatile functions, \eg, the Spring Framework, JUnit, or Google Guava. However, inaccurate or late support for these libraries may lead to inaccurate analysis results.
This characteristic is mainly related with the root cause of \emph{unhandled java libraries} (Section~\ref{rtcs:unhandled_java_libraries}).

\vspace{2pt}
\noindent\textbf{Variable Initialization and Assignments.}
Variable initializations and assignments may induce FPs or FNs.
This characteristic is mainly related with the root causes of \emph{missing specific cases} (Section~\ref{rtcs:missingcases}) and \emph{mishandling intermediate representations} (Section~\ref{rtcs:ir}).
For example, direct initialization (variables are assigned during initialization, \eg, {\small\texttt{int a = 1;}}), delayed assignment (variables are assigned after declaration, \eg, {\small\texttt{int a; ...; a = 1;}}), assignment within expressions (the assignments are within other expressions, \eg,  {\small\texttt{if(condition = var == 3)}}), nontrivial assignments (assignments involving other variables instead of only literals, \eg, {\small\texttt{a = b}}), and uninitialized variables can induce FPs or FNs.
Figure~\ref{fig:characteristics}f shows a FP of \textsc{PMD}'s rule \textit{UnusedAssignment}. This rule warns unused assignments and variables. The variable {\small\texttt{start}} is declared (line 2) and used (lines 3 and 4). However, the rule fails to handle the delayed variable assignment and incorrectly warns {\small\texttt{start}} is not used (line 2).

\vspace{2pt}
\noindent\textbf{Lambda Expressions.}
Lambda expression was introduced in Java 8. Failing to handle lambda expressions may induce issues.
This characteristic is mainly related with the root causes of \emph{unhandled language features} (Section~\ref{sec:unhandled_language_features}).
For example, Figure~\ref{fig:characteristics}g illustrates a FP of \textsc{PMD}'s rule \textit{UseDiamondOperator} induced by a lambda expression. This rule avoids the duplicate declarations of type parameters in the diamond operator. At line 2, the variable {\small\texttt{foo}} does not get explicitly typed, so the type declaration {\small\texttt{string}} in the diamond operator is not duplicated. However, this rule gives a FP only inside the lambda expression.

\vspace{2pt}
\noindent\textbf{Modifiers.}
Java modifiers (\eg, {\small\texttt{public}}, {\small\texttt{private}},  {\small\texttt{protected}}, {\small\texttt{static}} and {\small\texttt{final}}) control the accessibility of classes, constructors, fields or attributes. Analyzers may fail to deal with these modifiers.
Figure~\ref{fig:characteristics}h shows a FN of \textsc{SpotBugs} rule {\small\textit{ES\_COMPARING\_STRINGS\_WITH\_EQ}} triggered by the {\small\texttt{final}} modifier. This rule warns when the strings are compared with {\small\texttt{==}}. In this case, if {\small\texttt{s2}} is annotated with {\small\texttt{final}}, no warning is reported at line 5, although there is a comparison with {\small\texttt{==}}, which is a FN. Deleting the {\small\texttt{final}} modifier makes the rule work correctly.
This characteristic is mainly related with the root causes of \emph{unhandled language features} (Section~\ref{sec:unhandled_language_features}).

\vspace{2pt}
\noindent\textbf{Others.}
We find that a few minor characteristics (\eg, {\small\texttt{extend}},   {\small\texttt{enum}}, anonymous classes) that may also induce the issues of FNs/FPs. 

%% file: Sections/implication_and_discussion.tex
\section{Implications and Discussions}
\label{sec:implication}
This section discusses the implications distilled from the findings of \textbf{RQ1} and \textbf{RQ2} to shed light on what the developers and researchers could do to tackle FNs/FPs.
We will also discuss other aspects of our work.

\vspace{3pt}
\noindent\textbf{\emph{Avoid issues caused by common root causes or input characteristics}.}
Our study finds that \emph{unhandled language features}
is one of the most common root causes for \emph{all} the three studied analyzers (\textbf{RQ1}'s Finding 2). Thus, developers should \emph{timely} check the rule specifications when some new Java language features are introduced, and update the rule implementations if necessary.
According to the statistics in Table~\ref{tab:rootcause_categories}, timely supporting new language features could avoid 21.1\% of the issues.
\emph{Missing cases} is another common root cause (\textbf{RQ1}'s Finding 3).
To counter this, given a rule, the analyzers' developers should consider different input characteristics (summarized \textbf{RQ2}) when designing its test programs.
The code under check may be written in different specific ways.
In this regard, researchers could devise automated testing techniques to generate diverse test programs.
To show the feasibility, in Section~\ref{sec:transformation}, we designed such a proof-of-concept testing strategy to generate new test programs by mutating existing ones.
On the other hand, \textit{annotations}, \textit{method calls and field accesses}, and \textit{Java standard libraries} are the major input characteristics affecting all the three studied analyzers (see \textbf{RQ2}). Thus, developers should pay more attention to
model the semantics of annotations and Java standard libraries when implementing the rules.
According to the statistics in Figure~\ref{fig:proportion}, properly modeling \textit{annotations} could avoid 21\% of the issues.

\vspace{2pt}
\noindent\textbf{\emph{Improving the underlying static analysis modules}.}
Static code analyzers usually adopt some form of static analysis, \eg, AST-based syntactic pattern matching, data-flow analysis and symbolic execution. We find that \emph{improving the abilities of static analysis modules} in general is important for mitigating FNs/FPs. For example, \textsc{PMD} is mainly affected by \emph{type resolution error or limitation} and \emph{scope analysis error or limitation}  (\textbf{RQ1}'s Finding 4). As a result, \textsc{PMD} may incur FNs/FPs when handling \textit{same identifiers} (revealed by \textbf{RQ2}) which requires proper type resolution or scope analysis. Indeed, we note that \textsc{PMD} has recently significantly rewritten its type resolution for the new version v7.0. \textsc{SonarQube} may incur FNs/FPs due to the errors or limitations of its symbolic execution engine.
The developers of \textsc{SonarQube} are also continuing the improvement of the engine.

\rA{However, we observe that in some cases developers may not immediately improve the static analysis modules, although they would confirm the reported FNs/FPs.
    This is because they need to trade off many factors like the investment cost, the development plans, the tool performance and the analysis precision (or soundness) after the improvement.
    For example, \textsc{SonarQube}'s Issue \#90849~\cite{sonar_tickets} is an FN caused by the limitation of the symbolic execution engine. The developers confirmed this FN but do not plan to fix it immediately considering too much investment is required. The developers disclosed that the current engine is purposely designed in this way to avoid raising FPs at the cost of inducing FNs. But the developers commented that ``\emph{we are working on a new bug-detection engine, that should be, at some point, able to handle such cases and would allow us to replace this rule with a more performant version of it (this new engine is already running on \textsc{SonarCloud} and some versions of \textsc{SonarQube})''}.
    For another example, \textsc{PMD}'s developers confirmed several issues like \#4127~\cite{pmd4127} (which is a FP) which were caused by the limitations of type resolution. But it took the \textsc{PMD}'s developers more than one year to improve the type resolution module and resolve all these issues.

    Additionally, we observe that \emph{choosing the appropriate form of static analysis for implementing the rules} is also important. For example, data-flow analysis can provide more precise information than AST-based syntactic pattern matching, thus reducing potential FNs/FPs.
    In Section~\ref{sec:analyzer_weaknesses}, we will investigate the weaknesses of the static analysis modules in the studied analyzers, and show the consequences and trade-offs of choosing the forms of static analysis.
}

\vspace{2pt}
\noindent\textbf{\emph{Following best practices when building static code analyzers}.}
During our study, we observed some best practices to tackle FNs/FPs by inspecting the fixing patches. We explicitly summarize these best practices to inspire (new) developers.
(1) \emph{Enforcing modularity when designing rules}.
Some rules may have similar analysis procedures. In such scenarios, developers should consider moving these procedures into a common utility class.
In this way, fixing the FNs or FPs induced by this utility class could benefit all the relevant rules (no separated fixes are needed anymore). In the \textsc{PMD}'s PR \#2899, one developer commented ``\emph{using a common utility class to share logic between rules, or to store procedures that are not really rule-specific}''.
\textsc{PMD}'s developers commented in its roadmap ``\emph{In general, a rule should use TR (type resolution) when it can, and fall back on non-TR approach otherwise. No need for separating rules for TR and non-TR.}''~\cite{roadmap}.
These comments confirm the importance of modularity design.
(2) {\emph{Avoiding workaround fixes of FNs or FPs}.
The workaround fixes may introduce some adverse side effects --- fixing a FP but introducing new FNs, or visa versa.
Thus, developers should always try to fix the true root causes of FNs/FPs, and
carefully evaluate the effect of their patches.
During fixing, avoiding introducing FPs is also important because the number of reported FPs could affect the usability of the analyzers~\cite{google}.
For example, \textsc{SonarQube}'s developers commented that ``\emph{\textsc{SonarQube} is not warning against any good practice, rather the goal of each rule is to favor good practices while keeping the false positive rates as low as possible (ideally zero)}''~\cite{buono2023}.
Researchers could devise effective techniques to help developers find or avoid FPs.

%% file: Sections/Experiment.tex
\section{RQ3: Proof-of-Concept Demonstration of Our Findings}
\label{sec:proof_of_concept}
\rC{This section demonstrates how our findings can help identify issues in analyzers and reveal the weaknesses of static analysis modules.}
\subsection{\rA{Finding the Analyzers' FNs/FPs by Automatically Generating Equivalent Programs}}
\label{sec:transformation}

This section shows a proof-of-concept testing strategy to help find FNs or FPs of static code analyzers. The main idea is to automatically generate equivalent input programs from existing ones to stress test the analyzers.

\begin{table}[t]
	\caption{Equivalent Input Program Mutation Operators}
	\vspace*{-1pc}
	\label{tab:transformations}
	\centerline{\includegraphics[scale=0.8]{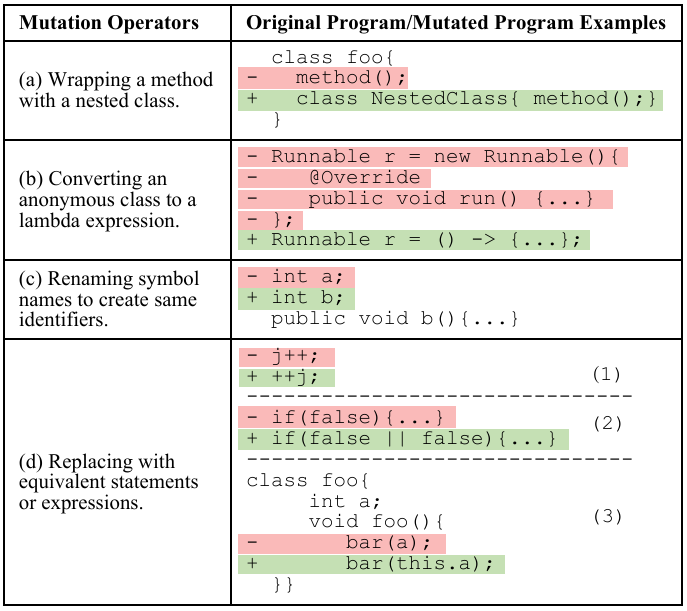}}
\end{table}

\vspace{2pt}
\noindent\textbf{Equivalent Input Program Mutation}.
One insight obtained from our study is that we can perform equivalent input program mutations to help find FNs or FPs.
\rB{Specifically, given an original input program of an analyzer, we can generate some equivalent program variants based on some program mutation operators. Here, equivalent program variants means that these variants are expected to have the same analysis results with the original program when checked by the analyzer. If the analysis results (\eg, the reported warnings) are different between the original program and its variants, some FNs or FPs are likely found.}
Based on the findings of \textbf{RQ2}, we selected four kinds of input characteristics to design the program mutation operators (see Table~\ref{tab:transformations}).
Specifically, based on the findings of \textbf{RQ1}, the mutation operators (a) and (b) target
the root cause of \emph{unhandled language features}, (c) targets the root cause of \emph{type resolution and scope analysis error or limitation}, and (d) targets the root cause of \emph{missing specific cases}.
\begin{enumerate}[leftmargin=*,label=(\alph*)]
	\item \emph{Wrapping a method with a nested class}.
	      Given a Java class, this mutation operator identifies all the methods in this class.
	      For each (\emph{static} or \emph{non-static}) method, the operator creates a nested class to wrap this method and properly adjusts the original class fields or methods accesses to be compliant with the code changes.

	\item \emph{Converting an anonymous class to a lambda expression}.
	      In Java, an anonymous class can be equivalently represented by a lambda expression. Thus, given a Java class, this mutation operator identifies all the anonymous classes. For each anonymous class, it converts the class to a lambda expression.

	\item \emph{Creating same identifiers}.
	      Given a Java class, this mutation operator identifies all the methods and the fields in this class. For each field, it renames its symbol name to the name of one existing method's name.

	\item \emph{Replacing with equivalent statements or expressions}. We support three forms of equivalent mutations: replacing (1) the statement {\small\texttt{j++;}} with {\small\texttt{++j;}} or vice versa, (2) {\small\texttt{false}} with {\small\texttt{false || false}}; and (3) the access of a non-static class field {\small\texttt{a}} with {\small\texttt{this.a}}.

\end{enumerate}

\vspace{2pt}
\noindent\textbf{Implementation}.
We use \textsc{JavaParser}~\cite{javaparser} to parse the input programs into abstract syntax trees (ASTs), manipulate the tree nodes according to the mutation operators and generate equivalent program variants.
The implementation consists of around 1000 lines of Java code.

\vspace{2pt}
\noindent\textbf{Experimental Setup}.
We applied our testing strategy to test the latest versions of the three studied analyzers at the time of our study (\textsc{PMD} v7.0.0-rc3, \textsc{Spotbugs} v4.7.4 and \textsc{SonarQube} v10.0.0.68432). We collected all the test programs released by these tools as the original input programs: 3,754 programs from \textsc{PMD}, 572 programs from \textsc{SonarQube}, and 1,227 programs from \textsc{Spotbugs}.
We used \textsc{JavaParser} to parse, transform and create new test programs in Java 11. We used Python scripts to run the tests against the analyzers and analyze the outputs. We compared the analysis outputs by checking whether the numbers or the types of reported warnings are identical. If not, we likely find some FNs/FPs, and manually inspect each of them for confirmation. 
The testing process was conducted on a 64-bit Ubuntu 20.04 LTS machine with 16GB RAM.
It took about 2 hours to run all the newly generated test programs for \textsc{PMD}, 1.5 hours for \textsc{SonarQube}, and 1 hour for \textsc{Spotbugs}.

\begin{table}[t]
	\caption{Statistics of the 19 issues found by our study}
	\vspace*{-1pc}
	\label{tab:issuetable}
	\centerline{\includegraphics[scale=0.8]{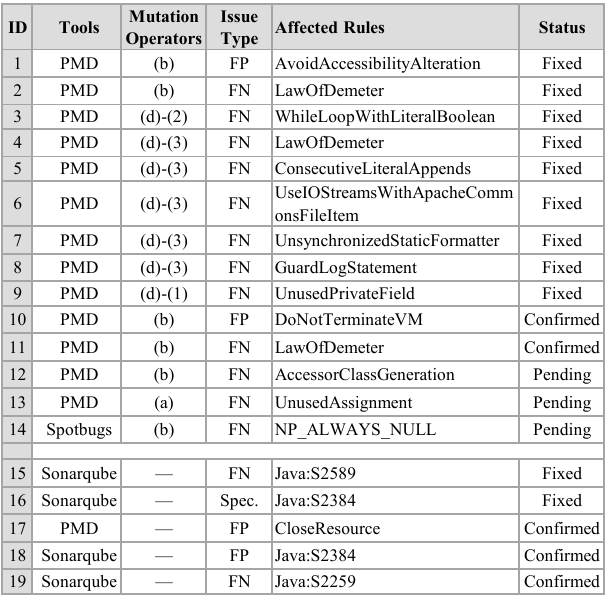}}
\end{table}

\vspace{2pt}
\noindent\textbf{Results and Analysis}.
Table~\ref{tab:issuetable} gives the issues of FNs and FPs (with Issue IDs from 1$\sim$14) found by our testing strategy. We found 12 FNs and 2 FPs from the three static code analyzers.
We reported all these issues to the developers. Up to now,
11 issues have been confirmed, 9 of which have been already fixed; and 3 issues
are still waiting for feedback from the developers.
These issues affected 12 different rules from \textsc{PMD} and \textsc{SpotBugs}.
We note that most issues were found by the mutation operator (b), which found 6 issues, and (d) (the three forms of equivalent expression) found 1, 1 and 5 issues, respectively.
The mutation (c) did not find new issues in the analyzers.
The reason may be that the latest version of \textsc{PMD} v7.0 significantly improved its type resolution and scope analysis modules. Thus, \textsc{PMD} may avoid many potential type resolution issues, while the other two analyzers are robust.

To our knowledge, Wang~\etal~\cite{wang2022find} conducted the first work to find FNs or FPs of static code analyzers.
They used a differential testing strategy, \ie, comparing the outputs of similar rules between two different analyzers to find issues.
Different from their work, our testing strategy is one form of metamorphic testing~\cite{chen1998metamorphic} which does not require a reference analyzer.
Indeed, 10 issues found by us which cannot be found by~\cite{wang2022find}.
Because the related 8 buggy rules in our experiment do not have similar rules in other analyzers. For example, for the rule \textit{``UseIOStreamsWithApacheCommonsFileItem''} in \textsc{PMD}, no similar rule in \textsc{SonarQube} exists.
Thus, the issue (with ID: 6 in Table~\ref{tab:issuetable})
could be missed by differential testing.
Thus, our proof-of-concept testing strategy could complement the prior work.
We believe that more mutation operators could be designed to help find more FNs or FPs of static code analyzers. We leave it as an interesting future work.

\subsection{\rA{Investigating the Weaknesses of Static Analysis Modules}}
\label{sec:analyzer_weaknesses}

The static code analyzers use some forms of static analysis.
Thus, the weaknesses of the static analysis modules in the analyzers may lead to FNs/FPs (see \textbf{RQ1}'s Finding 4).
To this end, we aim to manually investigate some typical rules of the studied analyzers to investigate their potential weaknesses based on our insights from \textbf{RQ1} and \textbf{RQ2}.

\vspace{2pt}
\noindent\textbf{Investigation Method}. To select typical rules for inspection,
we ranked all the rules of each studied analyzer in terms of the number of historical fixed issues from the most to the least.
To constraint our manual cost, we chose the top five buggy rules from \textsc{PMD}, \textsc{SpotBugs} and \textsc{SonarQube} for careful inspection.
Our insight is that investigating such ``buggy'' rules are more likely to reveal the weaknesses of static analysis modules. Table~\ref{tab:toprules} lists these selected rules from the three analyzers. In the column of ``Rule Name (\#Fixed Issues)'', the number in the parenthesis following the rule name is the number of historical fixed issues. The column ``Reference'' gives the rule specification.

\begin{table}[t]
	\footnotesize
	\centering
	\caption{Top five buggy rules of the studied analyzers in terms of the number of historical fixed issues.}
	\vspace*{-1pc}
	\label{tab:toprules}
	\begin{tabular}{|l|p{8cm}|c|}
		\hline
		\textbf{Tool}      & \textbf{Rule Name (\#Fixed Issues)}                                                        & \textbf{Reference}                   \\
		\hline
		\textsc{PMD}       & ImmutableField (12)                                                                        & \cite{ImmutableField}                \\
		                   & UnnecessaryFullyQualifiedName (11)                                                         & \cite{UnnecessaryFullyQualifiedName} \\
		                   & CloseResource (11)                                                                         & \cite{CloseResource}                 \\
		                   & UnusedPrivateField (8)                                                                     & \cite{UnusedPrivateField}            \\
		                   & UnusedImports (8)                                                                          & \cite{UnusedImports}                 \\
		\hline
		\textsc{SpotBugs}  & DMI\_RANDOM\_USED\_ONLY\_ONCE (2)                                                          & \cite{DMI}                           \\
		                   & NP\_NONNULL\_PARAM\_VIOLATION (2)                                                          & \cite{NP}                            \\
		                   & RV\_RETURN\_VALUE\_IGNORED (2)                                                             & \cite{RV}                            \\
		                   & RCN\_REDUNDANT\_NULLCHECK\_  OF\_NONNULL\_VALUE (2)                                        & \cite{RCN}                           \\
		                   & UPM\_UNCALLED\_PRIVATE\_METHOD (2)                                                         & \cite{UPM}                           \\
		\hline
		\textsc{SonarQube} & S2589: Boolean expressions should not be gratuitous (3)                                    & \cite{s2589}                         \\
		                   & S3749: Members of Spring components should be injected (3)                                 & \cite{s3749}                         \\
		                   & S2384: Mutable collection or array members should not be stored or returned directly (3)   & \cite{s2384}                         \\
		                   & S2259: Null pointers should not be dereferenced (3)                                        & \cite{s2259}                         \\
		                   & S2695: "PreparedStatement" and "ResultSet" methods should be called with valid indices (2) & \cite{s2695}                         \\
		\hline
	\end{tabular}
\end{table}

During our investigation, for each rule in Table~\ref{tab:toprules}, we (1) reviewed its specification and implementation (understanding which kind of program flaws the rule checks and which form of static analysis the rule uses),
(2) examined all its historical issues and the corresponding fixing patches
(understanding which root causes and/or input characteristics trigger these issues, and whether the corresponding fixing patches are workarounds), and
(3) applied some code mutations on the rule's test programs to manifest FNs/FPs.
Here, the code mutations are applied based on the findings of \textbf{RQ1} and \textbf{RQ2}: (i) the common input characteristics leading to FNs/FPs.
\rA{
From the top ten input characteristics triggering FPs/FNs identified in Section~\ref{sec:characteristic} (also see Fig.~\ref{fig:characteristics}), we selected
these three input characteristics, \ie, \textit{method calls and field accesses}, \textit{variable initialization and assignment} and \textit{complex expressions and statements}, as the mutation strategies (operators) to manually investigate the weaknesses of the analyzers. 
We did not select the other input characteristics from the top ten because some input characteristics (\eg, \textit{nested classes}, \textit{same identifiers}) are suitable for automatically generating equivalent programs (in Section~\ref{sec:transformation}) while some input characteristics (like \textit{annotations}, \textit{Java standard libraries} and \textit{Java external libraries}) are difficult to apply for manual code mutations.
In practice, we applied the input characteristic of \textit{method calls and field accesses} to test those rules which may require dataflow information (\eg, \textsc{PMD}'s rule \textit{CloseResource}). For example, we can encapsulate the sink (\eg, closing some resource) into a new method, and call that new method at the original place. This mutation strategy can stress test the inter-procedure analysis ability of a rule.
For another example, if the original test program involves some variable assignments, we can transform the assignments into different forms (\eg, transforming the assignment of a local variable into that of a global variable or a class variable) by applying the input characteristic of \textit{variable initialization and assignment} to stress test the rules.
We can also apply the characteristic of \textit{complex expressions and statements} to purposely transform the original expressions or statements into the complicated ones to stress test the rules.}
(ii) The correlations between the root causes and the input characteristics.
For example, mutating \emph{method calls and field accesses} may affect or require \emph{type resolution} and \emph{control/data-flow analysis}, and mutating \emph{variable initialization and assignments} and \textit{complex expressions or statements} may affect or require \emph{scope analysis}. Note that these code mutations do not guarantee the mutated program is equivalent to the rule's original test program (like what we did in Section~\ref{sec:transformation}).

Table~\ref{tab:issuetable} shows the five issues (with Issue IDs from 15$\sim$19) found by us in these analyzers' latest versions, indicating the weaknesses of their static analysis modules (leading to FNs or FPs). All these issues were confirmed by the developers. We illustrate these issues below.

\begin{figure*}[t]
	
	\includegraphics[width=\textwidth]{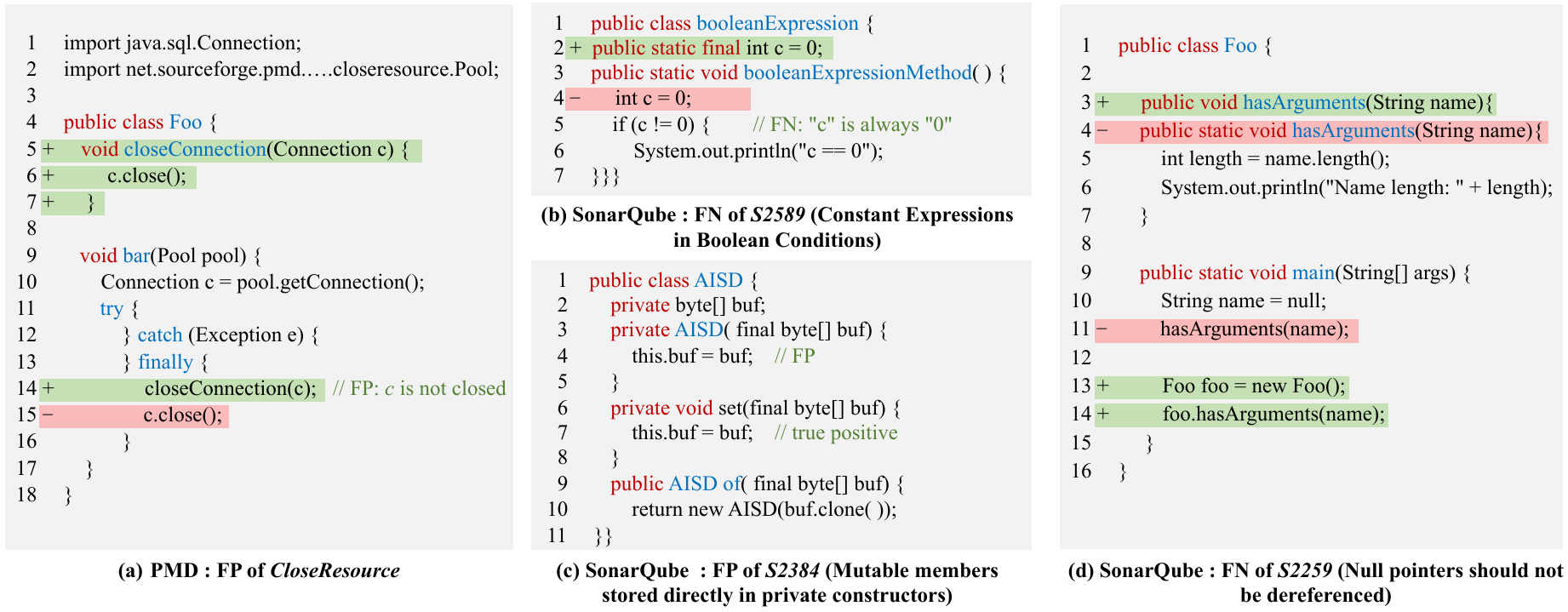}
	\centering
	\vspace*{-2pc}
	\caption{Simplified code examples illustrating the weaknesses of static analysis modules in the studied analyzers.}
	\label{fig:weaknesses}
\end{figure*}

\vspace{2pt}
\noindent\textbf{\emph{Fail to handle basic method call flows}}. Figure~\ref{fig:weaknesses}a shows a FP of \textsc{PMD}'s rule \textit{CloseResource}. In this case, we hoisted the original statement of closing the connection {\small\texttt{c}} (line 15) into a method call {\small\texttt{closeConnection}} (line 14), and closed the connection {\small\texttt{c}} at line 6. 
However, \textsc{PMD} reports a FP at line 14, although {\small\texttt{c}} is correctly closed at line 6.
The \textsc{PMD} developer confirmed that this issue is a valid FP and commented that ``\emph{since this is all within one class, it would be nice if \textsc{PMD} could detect this on its own (through some basic call flow)}''.
This case shows the weakness of data-flow analysis in \textsc{PMD}. In fact, we find that \textsc{PMD}'s data-flow analysis is limited (i.e., it only supports reaching definition analysis procedure~\cite{dataflowpass}).

\vspace{2pt}
\noindent\textbf{\emph{Fail to analyze variable scopes}}. Figure~\ref{fig:weaknesses}b shows a FN of \textsc{SonarQube}'s rule \textit{S2589}. This rule enforces that boolean expressions should not be gratuitous (if a boolean expression does not change the evaluation of the condition, it is redundant and should be removed).
In this example, we moved the declaration of the local variable {\small\texttt{c}} (line 4) outside of the method {\small\texttt{booleanExpressionMethod}} and declared
	{\small\texttt{c}} as a static final class field (line 2).
However, \textsc{SonarQube} failed to report the rule violation. We reported this issue to \textsc{SonarQube}. The developer confirmed that this is a valid FN and commented that ``\emph{Indeed our engine is failing to evaluate constants outside the method’s scope}''. This issue was fixed~\cite{2589_fixed}.
This case shows \textsc{SonarQube}'s weakness in analyzing variable scope.

\vspace{2pt}
\noindent\textbf{\emph{Fail to choose the appropriate form of static analysis}}.
Figure~\ref{fig:weaknesses}c shows a FP of \textsc{SonarQube}'s rule \textit{S2384}. This rule enforces that private mutable members in a class should not be directly stored or returned. In this example, this rule correctly warns line 7 because the private mutable member {\small\texttt{buf}} is directly stored.
However, the rule reports a FP at line 4 which should not be reported because the private constructor {\small\texttt{AISD}} is only called by the public method {\small\texttt{of}} which safely clones the parameter {\small\texttt{buf}}.
The \textsc{SonarQube} developer confirmed that this issue is a valid FP and commented that ``\emph{Unfortunately, this rule is (implemented as) AST-based, and it brings some limitations. So to eliminate these false positives the rule should rely on the data flow}''.
This case shows the weakness of \textsc{SonarQube} in choosing the inappropriate form of static analysis for some rules. We also found a flawed specification issue in this rule (Issue 16 in Table~\ref{tab:issuetable}) which has been fixed~\cite{2384_fixed}.

\vspace{2pt}
\noindent\textbf{\emph{Fail to track runtime types in symbolic execution}}.
Figure~\ref{fig:weaknesses}d shows a FN of \textsc{SonarQube}'s rule \textit{S2259}. This rule warns about null pointer dereference. For the original input program, this rule can correctly warn the dereferenced null pointer {\small\texttt{name}} (at line 5) because the string variable {\small\texttt{name}} is assigned as {\small\texttt{null}} (line 10).
However, when we changed the static method {\small\texttt{hasArguments}} to an instance method, and changed the original static method call on {\small\texttt{hasArguments}} to the method call by the class instance {\small\texttt{foo}}, the rule has a FN at line 5.
The \textsc{SonarQube} developer confirmed that this issue is a valid FN and commented that ``\emph{the FN is caused by a limitation of the symbolic execution engine}''.
The developer explained that, in the new code, the method {\small\texttt{hasArguments}} is an instance method and therefore requires a class instance to be called. The engine that runs this rule fails to track runtime types in the execution paths that it follows because it
currently only support tracking of final methods (\eg, static methods). The developer said they have already started to enhance the engine.
This case shows the weakness in resolving runtime types, affecting the precision of symbolic execution in \textsc{SonarQube}.

%% file: Sections/threats_to_validity.tex
\section{THREATS TO VALIDITY}

\noindent\textbf{Internal Validity.} \rC{Our study requires manual analysis and the human expertise of static code analyzers to answer RQ1 and RQ2. Thus, we may introduce some threats to the categories of root causes and input characteristics and their percentages. 
To counter this, in Sections~\ref{sec:rootcause} and \ref{sec:characteristic}, two of the co-authors independently inspected the issues, cross-checked their results, and discussed them with the other co-authors to reach a consensus. They tried their best to reduce potential threats.}

We studied 350 issues which include 270 FPs and 80 FNs (the FPs are more than the FNs). This disparity between FNs and FPs may bring some threats in the computed percentages of the categories of root causes and input characteristics.
But the readers should know that this disparity might be difficult to correct and it does not mean that FPs are more important than FNs. Because
(1) the users are more likely to report FPs than FNs, and (2) many FNs
are not reported because no checking rules exist~\cite{towhatextent,howmanyof,LiuCFLLXNLC23,LiCFFLLLC23} (such FNs may not be considered valid issues of the analyzers).

\noindent\textbf{External Validity.}
\rB{We collected and investigated 350 issues of FNs and FPs. The number of these issues might be not large enough and may bring some threats to the generability of our study's findings.
But we believe the generability of our findings can be justified by the following reason.
We have collected \emph{all} the historical, fixed issues of FNs and FPs from the three representative \textsc{PMD}, \textsc{Spotbugs} and \textsc{SonarQube} prior to the time of our study. These issues are diverse and were reported by the analyzers' own developers (who found FNs/FPs during development), real users (who found FNs/FPs by scanning real-world projects), and researchers (who found FNs/FPs by developing automated testing techniques).
Compared to the prior relevant work, our study investigates the largest number of issues of FNs and FPs and their patches.
For example, Thung \etal~\cite{towhatextent} and Habib \etal~\cite{howmanyof} only investigate 19 and 20 FNs respectively, Wang \etal~\cite{wang2022find} only investigate 46 issues (38 FNs and 8 FPs) and Zhang \etal~\cite{Statfier} only investigate 79 issues of FNs and FPs.
We have looked into these FNs/FPs which were analyzed by these prior work. We find our study's findings have included all their analysis results (\ie, root causes, input characteristics).}

\rA{One possible method of further generalizing some of our findings might be analyzing open-source Java projects by using these analyzers, and determine to what extent the coding practices in these projects are likely to ``trigger'' any of the issues we found.
However, this method may face some challenges in deciding the generalizability of our findings. First, it may be limited to investigating the issues of FPs if we do not have the ground-truth of program flaws in these projects. Second, if we do have the ground-truth, the missed flaws may not indicate the issues of FNs because the checking rules of these analyzers cannot cover all possible program flaws.
But we can access the generalizability of our findings from the perspective of Wang \etal's work~\cite{wang2022find}. They analyze 2,728 open-source Java projects by using these analyzers and find 46 issues of FNs/FPs based on differential testing. The root causes and input characteristics of these 46 issues are all included by our findings (which we will discuss in detail in Section~\ref{sec:related_work}), and thus our findings should be general.
}

In addition, \textsc{PMD} has more historical issues than the other two analyzers.
It may affect the generalizability of our findings.
But we find the root causes of PMD's issues are similar to those of \textsc{Spotbugs}'s and \textsc{SonarQube}'s issues. Thus, we believe the distilled categories of root causes should be general. \rB{Our study only considers three static code analyzers. So our conclusions may not generalize beyond these studied analyzers. However, these three analyzers are representative and widely used in practice and implement different forms of static analysis. In the future, to further mitigate the threats, we would expand our analysis to more static code analyzers.
We focus on the historical issues triggered by Java programs (Java is one of the most popular languages supported by existing static code analyzers~\cite{NachtigallSB22}). 
As a result, some of our findings may be specific to Java and may not be generalized for other programming languages.
Therefore, we plan to study the historical issues of FNs/FPs from other programming languages in the future to mitigate this potential threat.}

%% file: Sections/Relatedwork.tex
\section{RELATED WORK}
\label{sec:related_work}

This section discusses two strands of related work on studying static code analyzers: (1) evaluating the effectiveness and usability, and (2) finding and studying the FNs and FPs.

\vspace{2pt}
\noindent\textbf{Evaluating Static Code Analyzers}. \quad
In the literature, many studies exist in evaluating the effectiveness (\ie, the fault detection abilities) of the static code analyzers~\cite{effectiveness_2,effectiveness_3,towhatextent,towhatextent_journal,howmanyof,AreSonarQubeRules,groce2021evaluating,TomassiR21,LippBP22,MehrpourL23,LiuCFLLXNLC23,LiCFFLLLC23}.
All these studies reveal that static code analyzers suffer from FNs.
To analyze the reasons of FNs, for example, Thung \etal~\cite{towhatextent,towhatextent_journal} and Habib \etal~\cite{howmanyof} respectively manually examine 19 and 20 missed field defects and their corresponding program. They find that \emph{almost} all these defects were missed because they are not targeted by existing rules in the analyzers or they are domain-specific errors.
This insight is also shared by more recent and comprehensive studies~\cite{LiuCFLLXNLC23,LiCFFLLLC23} --- the insufficient rules \wrt field defects and the inability to handle logical (domain-specific) errors are the main reasons of FNs.
On the other hand, several studies reveal that static code analyzers are also affected by FPs~\cite{effectiveness_2,whydontsoftwared,ReynoldsJKPRH17,AreSonarQubeRules}, thus undermining the usability~\cite{NachtigallSB22,warnings_really_fixed,Howdevelopersengagewithstatic}.
Different from these prior studies, our work studies the analyzers from a new perspective, \ie, examining the historical (fixed) issues to understand FNs and FPs.
Moreover, our study inspects the implementations of static code analyzers, and the fixing patches to analyze FNs/FPs. Thus, many of our findings are fine-grained and have not been identified by these prior studies.
A recent comprehensive survey~\cite{mitigating_fp_survey} shows that, to mitigate FPs, most work develops post-processing techniques (\eg, statistical analysis, machine learning) to classify or rank the static analysis warnings. These work usually does not care about the implementation of static code analyzers. In contrast, our work inspects the implementations to find insights which could mitigate the FPs at the root.

\vspace{2pt}
\noindent\textbf{Finding and Studying FNs/FPs}. \quad
To our knowledge, the studies of Wang \etal~\cite{wang2022find} and Zhang \etal~\cite{Statfier}
are most relevant to ours. However, our study have several major differences with these two prior studies.
\emph{First}, the research goals of our study and these two studies are different.
Our work aims to conduct a systematic study on understanding the historical issues of FNs/FPs.
Thus, we investigate a broad range of 350 developer-confirmed \emph{and} -fixed FNs/FPs.
Wang \etal~\cite{wang2022find} and Zhang \etal~\cite{Statfier} mainly focus on designing some testing techniques to find FNs/FPs of the analyzers.
Although Wang \etal and Zhang \etal inspect the FNs/FPs \emph{found by their techniques} ---
Wang \etal inspect 46 found FNs/FPs (only 19 were confirmed \emph{and} fixed)
and Zhang \etal inspect 79 found FNs/FPs (only 26 were confirmed \emph{and} fixed) --- their conclusions could be biased by the limited diversity of their found issues.
Moreover, they do not examine the fixing patches and the implementations of the analyzers when  inspecting FNs/FPs.

\emph{Second}, due to the differences between research goals and the datasets,
our findings are more systematic and in-depth.
For example, the 13 ``bug patterns'' and 3 ``typical faults'' (see Section 3.2 and 3.3 in~\cite{wang2022find}) summarized by Wang \etal and 5 ``root causes'' (see Section 5.2 in~\cite{Statfier}) summarized by Zhang \etal are all included in our identified root causes and input characteristics.
Specifically, in Table~\ref{tab:rootcause_categories}, the root causes annotated by ``\textcolor{blue}{\halfcirc}'' and ``\textcolor{darkgreen}{\halfcirc}'' are only partially identified by Wang \etal and Zhang \etal respectively, while the root causes annotated by ``\textcolor{blue}{\emptycirc}'' and ``\textcolor{darkgreen}{\emptycirc}'' are missed by Wang \etal and Zhang \etal respectively.

\emph{Third}, Wang \etal use a differential testing strategy to find FNs/FPs, while we use a metamorphic testing strategy to find FNs/FPs (in Section~\ref{sec:transformation}).
\rA{These two testing strategies have their own strengths and weaknesses and can complement each other in finding FNs/FPs.
    The differential testing strategy might be limited by the number of paired rules with similar functionality between two different analyzers.
    Take \textsc{PMD} and \textsc{SonarQube} as an example, according to the statistics reported by Wang \etal (Section 3.1 in~\cite{wang2022find}), \textsc{PMD} and \textsc{SonarQube} respectively have 304 and 545 rules in total, but they only have 74 paired rules.
    Therefore, the differential testing strategy can only test 24.3\%($\approx$74/304) and 24.3\%($\approx$74/545) of all the rules of \textsc{PMD} and \textsc{SonarQube}, respectively, while the metamorphic testing strategy do not require the paired rules.
    On the other hand, the metamorphic testing strategy is limited to the number and strength of the identified metamorphic relations (\ie, the program mutation operators for generating equivalent program variants).
    In our case study, the issue finding results (discussed in Section~\ref{sec:transformation}) indeed show that our metamorphic testing strategy can complement the differential testing strategy because 10 out of the 14 FNs/FPs found by our metamorphic testing strategy cannot be found by the differential testing strategy.
    The reason is that the rules affected by these 10 FNs/FPs do not have the paired rules, and thus cannot be tested by differential testing.
    But our metamorphic testing strategy does not have such limitations.
    On the other hand, the rules affected by the remaining 4 FNs/FPs could be found by the differential testing strategy because these rules have their paired rules.
    It would be interesting in the future to further compare the differential testing strategy and our metamorphic testing strategy in a more systematic way.
    But in general it is difficult to decide which testing strategy is more effective in finding FNs/FPs because different factors like the diversity of input programs and the number/strength of identified metamorphic relations may affect the results.}
Zhang \etal use metamorphic testing like us, and propose 13 mutation operators (3 operators are inspired from historical issues).
However, only one mutation operator (see Table~\ref{tab:transformations}, (d)-(2)) used by our testing strategy is overlapped with those of Zhang \etal (see Table 2 in~\cite{Statfier}).
The remaining five mutation operators (see Table~\ref{tab:transformations}, (a), (b), (c), (d)-(1), (d)-(3)) from our work are new and have not been identified by Zhang \etal.
This difference also reflects that our study's findings are more systematic.
As a result, the testing technique proposed by Zhang \etal could find only one issue (\ie, Issue 3 in Table~\ref{tab:issuetable}) out of 14 FNs/FPs found by us.
Last but not least, our work identifies additional four FNs/FPs caused by four types of weaknesses of the static analysis modules, while Wang \etal and Zhang \etal do not perform such deep analysis on the static analysis modules.

In the literature, there are also other works focused on finding defects in static code analyzers. These works focus on either specific static analysis modules or specific input program characteristics, \eg, value analysis and constant propagation~\cite{cuoq2012testing}, alias analysis~\cite{alias_analysis_errors13}, data-flow analysis~\cite{taneja2020testing}, the configurations of static analysis~\cite{MordahlW21,ecstatic23}, and annotation-introduced faults~\cite{annotationIntroduced}.

There are some work validating the correctness of more sophisticated program analyzers like abstract interpreters~\cite{MidtgaardM15}, symbolic executors~\cite{kapus2017automatic}, model checkers~\cite{zhang2019finding} and compilers~\cite{le2014compiler}.
But these work does not target the analyzers we studied.

%% file: Sections/Conclusion.tex
\section{CONCLUSION}

We present the first systematic study on 350 historical issues of FNs/FPs from three representative static code analyzers. We investigated the root causes and input characteristics, which help  developers and researchers to understand FNs/FPs. Our study yields some new interesting findings and implications to improve the static code analyzers. Additionally, we conduct two proof-of-concept demonstrations to show the usefulness of our findings: (1) finding FNs and FPs, and (2) investigating the weaknesses of the static analysis modules. We have made our artifacts publicly available at \rC{\textit{\url{https://zenodo.org/doi/10.5281/zenodo.11525129
}}} to benefit the community.